\documentclass[twocolumn]{aastex631}
\usepackage{amsmath}
\usepackage{epsf}
\usepackage{color}
\usepackage{color}
\usepackage{graphicx}
\usepackage{color}
\usepackage{enumitem}
\usepackage{mathtools}
\usepackage[mathcal]{eucal}
\usepackage{tabulary}
 \let\mathscr\relax
\usepackage[scr]{rsfso}

\shorttitle{NGDEEP Survey Overview}
\shortauthors{Bagley et al.}
\newcommand{\sol}{$_{\odot}$}
\newcommand{\lya}{Ly$\alpha$}

\newcommand{\hst}{{\it HST}}

\def\arcs{\hbox{$^{\prime\prime}$}}

\def\W{$\lambda$}

\newcommand{\Ha}{\textrm{H}$\alpha$}
\newcommand{\Hb}{\textrm{H}$\beta$}

\newcommand{\MgII}{\ion{Mg}{2}}

\newcommand{\OII}{[\ion{O}{2}]}
\newcommand{\NeIII}{[\ion{Ne}{3}]}

\newcommand{\OIII}{[\ion{O}{3}]}

\newcommand{\SII}{[\ion{S}{2}]}

\newcommand{\ergs}{\ifmmode \textrm{erg\ s}^{-1} \else erg s$^{-1}$\fi}
\newcommand{\Msun}{\ifmmode \textrm{M}_{\odot} \else $M_{\odot}$\fi}
\newcommand{\Lsun}{\ifmmode \textrm{L}_{\odot} \else $L_{\odot}$\fi}

\newcommand{\HST}{\textit{HST}}
\newcommand{\JWST}{\textit{JWST}}

\definecolor{red}{rgb}{1,0,0}

\begin{document}
\title{The Next Generation Deep Extragalactic Exploratory Public
  (NGDEEP) Survey}

\author[0000-0002-9921-9218]{Micaela B. Bagley}
\affiliation{Department of Astronomy, The University of Texas at
  Austin, Austin, TX, USA}

\author[0000-0003-3382-5941]{Nor Pirzkal}
\affiliation{ESA/AURA}
\affiliation{Space Telescope Science Institute, 3700 San Martin Drive, Baltimore, MD 21218, USA}

\author[0000-0001-8519-1130]{Steven L. Finkelstein}
\affiliation{Department of Astronomy, The University of Texas at Austin, Austin, TX, USA}
\email{stevenf@astro.as.utexas.edu}

\author[0000-0001-7503-8482]{Casey Papovich}
\affiliation{Department of Physics and Astronomy, Texas A\&M University, College Station, TX, 77843-4242 USA}
\affiliation{George P.\ and Cynthia Woods Mitchell Institute for Fundamental Physics and Astronomy, Texas A\&M University, College Station, TX, 77843-4242 USA}

\author[0000-0002-4153-053X]{Danielle A. Berg}
\affiliation{Department of Astronomy, The University of Texas at Austin, Austin, TX, USA}

\author[0000-0003-3130-5643]{Jennifer M. Lotz}
\affiliation{Gemini Observatory/NSF's National Optical-Infrared Astronomy Research Laboratory, 950 N. Cherry Ave., Tucson, AZ 85719, USA}

\author[0000-0002-9393-6507]{Gene C. K. Leung}
\affiliation{Department of Astronomy, The University of Texas at Austin, Austin, TX, USA}

\author[0000-0001-7113-2738]{Henry C. Ferguson}
\affiliation{Space Telescope Science Institute, 3700 San Martin Drive, Baltimore, MD 21218, USA}

\author[0000-0002-6610-2048]{Anton M. Koekemoer}
\affiliation{Space Telescope Science Institute, 3700 San Martin Drive, Baltimore, MD 21218, USA}

\author[0000-0001-5414-5131]{Mark Dickinson}\affiliation{NSF's National Optical-Infrared Astronomy Research Laboratory, 950 N. Cherry Ave., Tucson, AZ 85719, USA}

\author[0000-0001-9187-3605]{Jeyhan S. Kartaltepe}
\affiliation{Laboratory for Multiwavelength Astrophysics, School of Physics and Astronomy, Rochester Institute of Technology, 84 Lomb Memorial Drive, Rochester, NY 14623, USA}\

\author[0000-0002-8360-3880]{Dale D. Kocevski}
\affiliation{Department of Physics and Astronomy, Colby College, Waterville, ME 04901, USA}

\author[0000-0002-6748-6821]{Rachel S. Somerville}
\affiliation{Center for Computational Astrophysics, Flatiron Institute, 162 5th Avenue, New York, NY, 10010, USA}

\author[0000-0003-3466-035X]{{L. Y. Aaron} {Yung}}
\altaffiliation{NASA Postdoctoral Fellow}
\affiliation{Astrophysics Science Division, NASA Goddard Space Flight Center, 8800 Greenbelt Rd, Greenbelt, MD 20771, USA}

\author[0000-0001-8534-7502]{Bren E. Backhaus }
\affil{Department of Physics, University of Connecticut, Storrs, CT 06269, USA}

\author[0000-0002-0930-6466]{Caitlin M. Casey}
\affiliation{Department of Astronomy, The University of Texas at Austin, Austin, TX, USA}
\affiliation{Cosmic Dawn Center (DAWN), Denmark}

\author[0000-0001-9875-8263]{Marco Castellano}
\affiliation{INAF - Osservatorio Astronomico di Roma, via di Frascati 33, 00078 Monte Porzio Catone, Italy}

\author[0000-0002-0786-7307]{\'{O}scar A. Ch\'{a}vez Ortiz}
\affiliation{Department of Astronomy, The University of Texas at Austin, Austin, TX, USA}

\author[0000-0003-4922-0613]{Katherine Chworowsky}
\affiliation{Department of Astronomy, The University of Texas at Austin, Austin, TX, USA}
\altaffiliation{NSF Graduate Fellow}

\author[0000-0002-1803-794X]{Isabella G. Cox}
\affiliation{Laboratory for Multiwavelength Astrophysics, School of Physics and Astronomy, Rochester Institute of Technology, 84 Lomb Memorial Drive, Rochester, NY 14623, USA}

\author[0000-0003-2842-9434]{Romeel Dav\'e}
\affiliation{Institute for Astronomy, University of Edinburgh, Royal Observatory, Edinburgh, EH9 3HJ, UK}
\affiliation{University of the Western Cape, Bellville, Cape Town 7535, South Africa}

\author[0000-0001-8047-8351]{Kelcey Davis}
\altaffiliation{NSF Graduate Research Fellow}
\affiliation{Department of Physics, 196 Auditorium Road, Unit 3046, University of Connecticut, Storrs, CT 06269, USA}

\author[0000-0001-8489-2349]{Vicente Estrada-Carpenter}\affiliation{Department of Astronomy \& Physics, Saint Mary's University, 923 Robie Street, Halifax, NS, B3H 3C3, Canada}

\author[0000-0003-3820-2823]{Adriano Fontana}
\affiliation{INAF Osservatorio Astronomico di Roma, Via Frascati 33, 00078 Monteporzio Catone, Rome, Italy}

\author[0000-0001-7201-5066]{Seiji Fujimoto}
\altaffiliation{Hubble Fellow}
\affiliation{Department of Astronomy, The University of Texas at Austin, Austin, TX, USA}

\author[0000-0003-2098-9568]{Jonathan P. Gardner}
\affiliation{Astrophysics Science Division, NASA Goddard Space Flight Center, 8800 Greenbelt Rd, Greenbelt, MD 20771, USA}

\author[0000-0002-7831-8751]{Mauro Giavalisco}
\affiliation{Department of Astronomy, University of Massachusetts, Amherst, MA 01003, USA}
\
\author[0000-0002-5688-0663]{Andrea Grazian}
\affil{INAF--Osservatorio Astronomico di Padova,
Vicolo dell'Osservatorio 5, I-35122, Padova, Italy\\}

\author[0000-0001-9440-8872]{Norman A. Grogin}
\affiliation{Space Telescope Science Institute, 3700 San Martin Drive, Baltimore, MD 21218, USA}

\author[0000-0001-6145-5090]{Nimish P. Hathi}
\affiliation{Space Telescope Science Institute, 3700 San Martin Drive, Baltimore, MD 21218, USA}

\author[0000-0001-6251-4988]{Taylor A. Hutchison}
\altaffiliation{NASA Postdoctoral Fellow}
\affiliation{Astrophysics Science Division, NASA Goddard Space Flight Center, 8800 Greenbelt Rd, Greenbelt, MD 20771, USA}

\author[0000-0002-6790-5125]{Anne E. Jaskot}
\affiliation{Department of Astronomy, Williams College, Williamstown, MA, USA}

\author[0000-0003-1187-4240]{Intae Jung}
\affiliation{Space Telescope Science Institute, 3700 San Martin Drive, Baltimore, MD 21218, USA}

\author[0000-0001-8152-3943]{Lisa J. Kewley}
\affiliation{Center for Astrophysics | Harvard \& Smithsonian, 60 Garden Street, Cambridge, MA 02138, USA}

\author[0000-0002-5537-8110]{Allison Kirkpatrick}
\affiliation{Department of Physics and Astronomy, University of Kansas, Lawrence, KS 66045, USA}

\author[0000-0003-2366-8858]{Rebecca L. Larson}
\altaffiliation{NSF Graduate Fellow}
\affiliation{Department of Astronomy, The University of Texas at Austin, Austin, TX, USA}

\author[0000-0002-7547-3385]{Jasleen Matharu}
\affiliation{Cosmic Dawn Center, Niels Bohr Institute, University of Copenhagen, R\aa dmandsgade 62, 2200 Copenhagen, Denmark\\}

\author[0000-0002-5554-8896]{Priyamvada Natarajan}
\affiliation{Department of Astronomy, Yale University, 52 Hillhouse Avenue, New Haven, CT 06511, USA}
\affiliation{Department of Physics, Yale University, P.O. Box 208121, New Haven, CT 06520, USA}
\affiliation{Black Hole Initiative at Harvard University, 20 Garden Street, Cambridge, MA 02138, USA}

\author[0000-0001-8940-6768]{Laura Pentericci}
\affiliation{INAF - Osservatorio Astronomico di Roma, via di Frascati 33, 00078 Monte Porzio Catone, Italy}

\author[0000-0003-4528-5639]{Pablo G. P\'erez-Gonz\'alez}
\affiliation{Centro de Astrobiolog\'{\i}a (CAB), CSIC-INTA, Ctra. de Ajalvir km 4, Torrej\'on de Ardoz, E-28850, Madrid, Spain}

\author[0000-0002-5269-6527]{Swara Ravindranath}
\affiliation{Space Telescope Science Institute, 3700 San Martin Drive, Baltimore, MD 21218, USA}

\author[0000-0003-2283-2185]{Barry Rothberg}
\affiliation{Department of Physics and Astronomy, George Mason University, 4400 University Drive, MSN 3F3, Fairfax, VA 22030, USA}
\affiliation{U.S. Naval Observatory, 3450 Massachusetts Avenue NW, Washington, DC 20392, USA}

\author[0000-0003-0894-1588]{Russell Ryan}
\affiliation{Space Telescope Science Institute, 3700 San Martin Drive, Baltimore, MD 21218, USA}

\author[0000-0001-9495-7759]{Lu Shen}
\affiliation{Department of Physics and Astronomy, Texas A\&M University, College Station, TX, 77843-4242 USA}
\affiliation{George P.\ and Cynthia Woods Mitchell Institute for Fundamental Physics and Astronomy, Texas A\&M University, College Station, TX, 77843-4242 USA}

\author[0000-0002-6386-7299]{Raymond C. Simons}
\affiliation{Department of Physics, 196 Auditorium Road, Unit 3046, University of Connecticut, Storrs, CT 06269, USA}

\author[0000-0002-4226-304X]{Gregory F. Snyder}
\affiliation{Space Telescope Science Institute, 3700 San Martin Drive, Baltimore, MD 21218, USA}

\author[0000-0002-1410-0470]{Jonathan R. Trump}
\affiliation{Department of Physics, 196 Auditorium Road, Unit 3046, University of Connecticut, Storrs, CT 06269, USA}

\author[0000-0003-3903-6935]{Stephen M.~Wilkins} %
\affiliation{Astronomy Centre, University of Sussex, Falmer, Brighton BN1 9QH, UK}
\affiliation{Institute of Space Sciences and Astronomy, University of Malta, Msida MSD 2080, Malta}

\begin{abstract}
We present the Next Generation Deep Extragalactic Exploratory Public (NGDEEP) Survey, a deep slitless spectroscopic and imaging Cycle 1 \JWST\ treasury survey designed to constrain feedback mechanisms 
in low-mass galaxies across cosmic time.  NGDEEP targets the Hubble Ultra Deep Field (HUDF) with NIRISS slitless spectroscopy (f$_{\mathrm{lim,line},5\sigma} \approx 1.2\times$10$^{-18}$ erg/s/cm$^2$) to measure metallicities and star-formation rates (SFRs) for low-mass galaxies through the peak of the cosmic SFR density ($0.5<z<4$).  In parallel, NGDEEP targets the HUDF-Par2 parallel field with NIRCam ($m_{\mathrm{lim},5\sigma}=30.6-30.9$) to discover galaxies to $z>12$, constraining the slope of the faint-end of the rest-ultraviolet luminosity function.  NGDEEP overlaps with the deepest {\it HST} ACS optical imaging in the sky: F435W in the HUDF ($m_{\mathrm{lim,F435W}}=29.6$), and F814W in HUDF-Par2 ($m_{\mathrm{lim,F814W}}=30$), making this a premier \HST+\JWST\ Deep Field. As a treasury survey,  NGDEEP data is public immediately, and we will rapidly release data products and catalogs in the spirit of previous deep field initiatives. In this paper we present the NGDEEP survey design, summarize the science goals, and detail plans for the public release of NGDEEP reduced data products. 
\end{abstract}


\section{Introduction}\label{sec:intro}

Deep field observations push astronomical source detection to the faintest accessible limits.  They are often motivated by the wish to discover, count, and study the most distant objects. Deep fields also survey the faintest objects detectable at intermediate distances, constraining luminosity functions and other statistical properties of the evolving galaxy population.

Although the \textit{Hubble Space Telescope} (\HST) Hubble Deep Field may be the most iconic early deep field, it was not the first such observation. Astronomers took deep images of ``blank'' high-latitude fields using photographic plates \citep[e.g.,][]{Kron1978,Koo1981} and CCDs \citep[e.g.,][]{Tyson1979} and at radio wavelengths \citep[e.g.,][]{Windhorst1985}.  Measuring and analyzing faint galaxy number counts was a popular pastime, motivated in part by cosmological goals, but also  providing evidence for the evolution of galaxies with cosmic time and distance.  Observations through two or more filters led to the recognition of the abundant population of ``faint blue galaxies'' as further evidence for an evolving galaxy population \citep[]{KooKron1992,Ellis1997}, and measurements with three or more filters were used to estimate galaxy redshifts \citep[]{Koo1985} and to identify very distant galaxy candidates via distinctive color signatures caused by the redshifted Lyman break \citep[]{Guhathakurta1990, Steidel1992}.  

The first Hubble Deep Field \citep[HDF,][]{Williams1996} was conceived as a public survey with non-proprietary data products available to any researcher.  The HDF data were indeed widely used by the community, and catalyzed extensive follow-up imaging and spectroscopy from ground- and space-based observatories, which further enriched the resources of widely-available data to study galaxy evolution. The HDF was later observed with \HST's near-infrared camera NICMOS \citep{Thompson1998,Thompson1999,Dickinson2000}, detecting redshifted optical rest frame light from galaxies out to $z \approx 3$ and extending the wavelength baseline for photometric redshift and spectral energy distribution (SED) analysis.  The installation of a more sensitive Advanced Camera for Surveys \citep[ACS;][]{clampin2000} during the second Hubble servicing mission motivated a Hubble Ultra Deep Field \citep[HUDF,][]{Beckwith2006}, with subsequent infrared follow-up with 
NICMOS \citep[]{Thompson2005}, the HUDF parallel program \citep[]{oesch07}, and later with the more sensitive WFC3 \citep{kimble08} infrared channel \citep[]{oesch10, Ellis2013, koekemoer13, illingworth13}.
The deep infrared data were used to identify and study galaxies with photometric redshifts as high as $z \approx 12$. 

Space observatories (e.g., Chandra, ISO, Spitzer, and Herschel) conducted their own deep field programs at X-ray and mid- to far-infrared wavelengths, typically in fields already surveyed by Hubble and ground-based facilities, including the HDF and the HUDF.  \textit{HST} itself revisited its deep fields many times, including deep observations using slitless spectroscopy with ACS and WFC3 to measure redshifts and other spectral properties for faint objects without spectroscopic pre-selection, including GRAPES 
and FIGS (PIDs 9793, 
13779, PI S. Malhotra; \citealt{pirzkal04,malhotra05,rhoads09,pirzkal17}).


The potential of \JWST\ deep fields was evident immediately in Cycle 1 with the Early Release Observation of the galaxy cluster and gravitational lens SMACS 0723.3-7327 \citep[PID 2736, PI: K. Pontoppidan;][]{pontoppidan22}. 
The GLASS-JWST Early Release Science Program \citep[PID 1324, PI: T. Treu;][]{treu22} observed the Abell 2744 lensing galaxy cluster, obtaining spectroscopy on the cluster with deep parallel imaging 
that reaches a 5$\sigma$ point source depth of $\sim$30.2 mag in $\sim$15.7 hours \citep{paris23}. 
The JADES (PID 1180 PI: D. Eisenstein, \citealt{prop1180}; 1210, 1286, 1287, PI: N. Luetzgendorf, \citealt{prop1210,prop1286,prop1287}) and MIRI Deep Survey \citep[PID 1283, PI: H. U. N\o{}rgaard-Nielsen and G. \"{O}stlin;][]{prop1283} GTO Programs are continuing the legacy of deep imaging and spectroscopy in and around the HUDF.  
Together the imaging portions of these early programs have easily detected galaxy candidates out to $z \approx 16$ \citep[e.g.,][]{adams22,castellano22,donnan22,harikane22,naidu22,robertson2022b,atek23,perez-gonzalez23}, while the spectroscopic portions of the programs are confirming galaxies from $z\sim7$ to 13 \citep[e.g.,][]{curtis-lake2022,roberts-borsani22,schaerer22}. These and other Cycle 1 surveys represent only the beginning of the deep field science made possible by \JWST's great leap in sensitivity and exquisite angular resolution.

Here we present the Next Generation Deep Extragalactic Exploratory Public (NGDEEP, PID 2079, PIs: S. Finkelstein, C. Papovich, N. Pirzkal) Survey\footnote{Originally named WDEEP in our Cycle 1 proposal}, which follows in the footsteps of previous treasury deep fields, probing a well-studied blank region of the sky, and providing data to the community with no proprietary period.
NGDEEP leverages {\it JWST}'s parallel observation capabilities to obtain \emph{two} deep fields for the price of one.  NGDEEP primary observations are comprised of deep NIRISS wide-field slitless spectroscopy \citep[WFSS;][]{doyon2012,willott2022} covering the HUDF proper \citep{beckwith06}.  At the same time, NGDEEP will obtain deep NIRCam imaging \citep{rieke2003,rieke2005,rieke22} in parallel covering the HUDF05-02 parallel field. We refer to these two observations as NGDEEP-NIS and NGDEEP-NRC, respectively.


\begin{figure*}
\epsscale{1.2}
\plotone{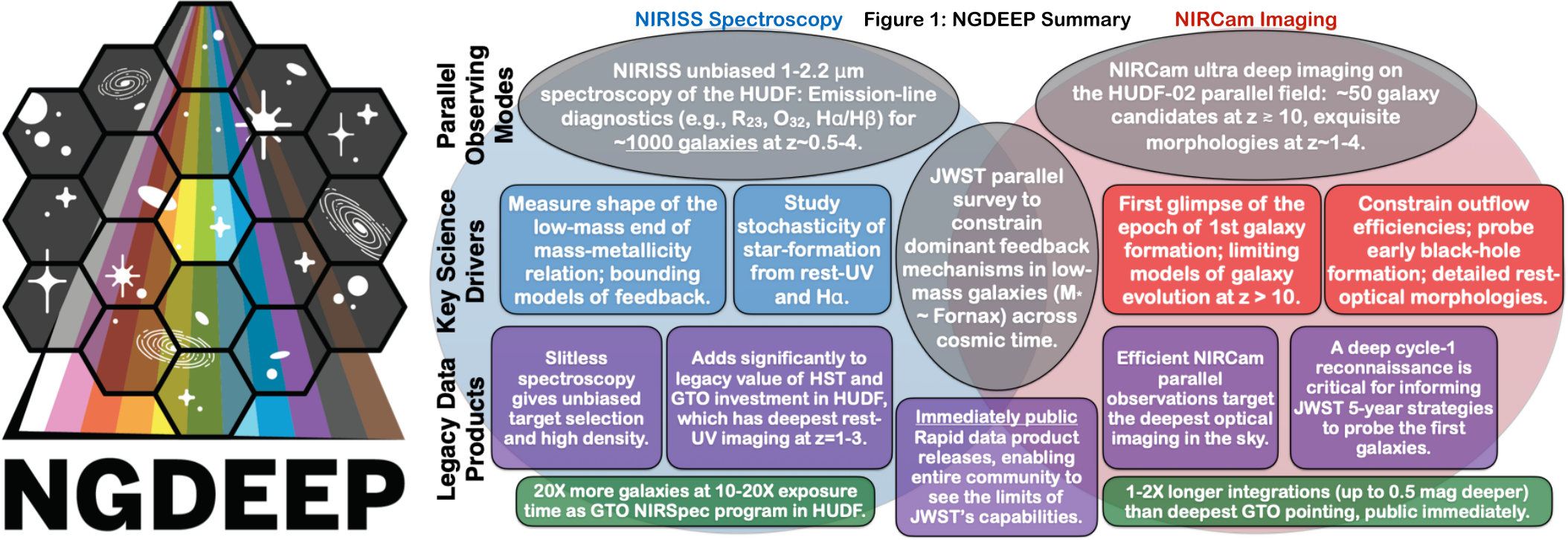}
\caption{\textit{Left:} The NGDEEP logo.  The hexagonal shapes represent \JWST’s mirror, with depictions of a variety of astrophysical objects, representing the variety of science possible with our survey.  The dispersion of colors represents both the spectroscopic component of NGDEEP, and also the NGDEEP team’s commitment to the inclusion of all members. \textit{Right:} A summary of the NGDEEP observing modes, key science drivers, and legacy data products.
\label{fig:barbados}}
\end{figure*}

NGDEEP is designed to illuminate the processes regulating galaxy evolution across cosmic time. 
Galaxy evolution is the result of a complex interplay between gas accretion from the cosmic web, the cooling and conversion of dense gas into stars, pollution of the interstellar medium (ISM) with heavy elements, and stellar-- and black--hole driven feedback processes which can disperse heavy elements into  the circumgalactic and intergalactic media (CGM, IGM). 
Feedback from massive stars and supernovae is a crucial element in regulating galaxy formation and shaping observable properties of galaxies, including their stellar mass, star formation history, gas fraction, metallicity, and morphological structure \citep[][and references therein]{somerville15}. Yet, the physical details of feedback remain highly uncertain.
  
Current cosmological simulations cannot calculate the feedback processes directly, thus they make assumptions which differ significantly from one simulation to another \citep[e.g.,][]{somerville15}.
Gaining insights into how feedback works requires constraining the observables that are the most sensitive to these divergent assumptions.  This task includes testing the inflows and outflows of heavy elements through observations of the mass--metallicity relation,  
measuring the amplitude and timescales of star-formation, and constraining 
star--formation efficiency in galaxies.

Together, both the primary NGDEEP-NIS and parallel NGDEEP-NRC observations will constrain the mechanisms regulating feedback in low--mass galaxies. 
At $z\sim1-3$, NGDEEP-NIS will obtain spectroscopy of the diagnostic emission lines in these $m_{\mathrm{UV}} \sim 28$~mag galaxies, leveraging the multiplexing of NIRISS WFSS to measure S/N~$>3$ (5) emission lines of over 2000 (1500) galaxies (see Table~\ref{tab:ncounts}). 
These observations will enable (1) measurements of multi-line metallicities for $>$350 galaxies at $z \sim$ 0.5--4 down to $\log(M_\ast/M_\odot){\sim}7$, testing predictions for the low-mass slope of the mass-metallicity relation (MZR) where models currently diverge \citep[e.g.,][]{ma16}; (2) measurements of H$\alpha$-based SFRs for $\sim$1000 low-mass galaxies at $0.7\lesssim z \lesssim2.3$ to $\sim$ 0.1 M\sol/yr, matching the UV SFR limit from the ultra-deep HUDF F435W imaging.  This will quantify the level of and constrain the effects of the stochasticity of SFRs in galaxies 100$\times$ lower in mass than the best NIR spectroscopic surveys \citep{kriek15,shivaei15}.

At $z > 9$ low--mass galaxies are predicted to be $\sim$30 mag and are only accessible by deep NIRCam imaging, thus NGDEEP-NRC will probe the population of faint galaxies ($M_{\mathrm{UV}} = -17.5$, $10\sigma$) out to the highest redshifts ($z\sim10-15$), constraining the faint (low-mass) end of the rest-UV  luminosity (stellar mass) function via the discovery of $\sim$ 100  galaxies at $z >$ 10, and up to 10 galaxies at $z \gtrsim$ 14.  This unique combination of NIRISS prime spectroscopy and NIRCAM parallel imaging will provide in-depth insight into galaxy evolution across cosmic time.

In this paper we present the NGDEEP survey, its design, scientific motivation, and pre-survey predictions.  In \S~\ref{sec:design} we present the survey design, field orientation and observing timeline, describing the NGDEEP NIRISS (NGDEEP-NIS) and NIRCam (NGDEEP-NRC) observations in \S~\ref{sec:nis_obs} and \S~\ref{sec:nrc_obs}, respectively. We discuss the leading science cases for NGDEEP-NIS in \S~\ref{sec:nis_science} and for NGDEEP-NRC in \S~\ref{sec:nrc_science}. As part of the survey design and verification process, we performed end-to-end simulations for both observations, which we present in \S~\ref{sec:simulations}. Finally, we outline our timeline for the public release of NGDEEP data products in \S~\ref{sec:datareleases} and briefly summarize in \S~\ref{sec:summary}.
We express all magnitudes in the AB system \citep{oke83} unless otherwise noted.


\section{Survey Design} \label{sec:design}

In this section, we describe the design of the NGDEEP survey and the capabilities of each set of observations. 
We provide a high-level summary of NGDEEP observations, science drivers, and planned legacy data products in Figure~\ref{fig:barbados}. 

\begin{figure*}
\plotone{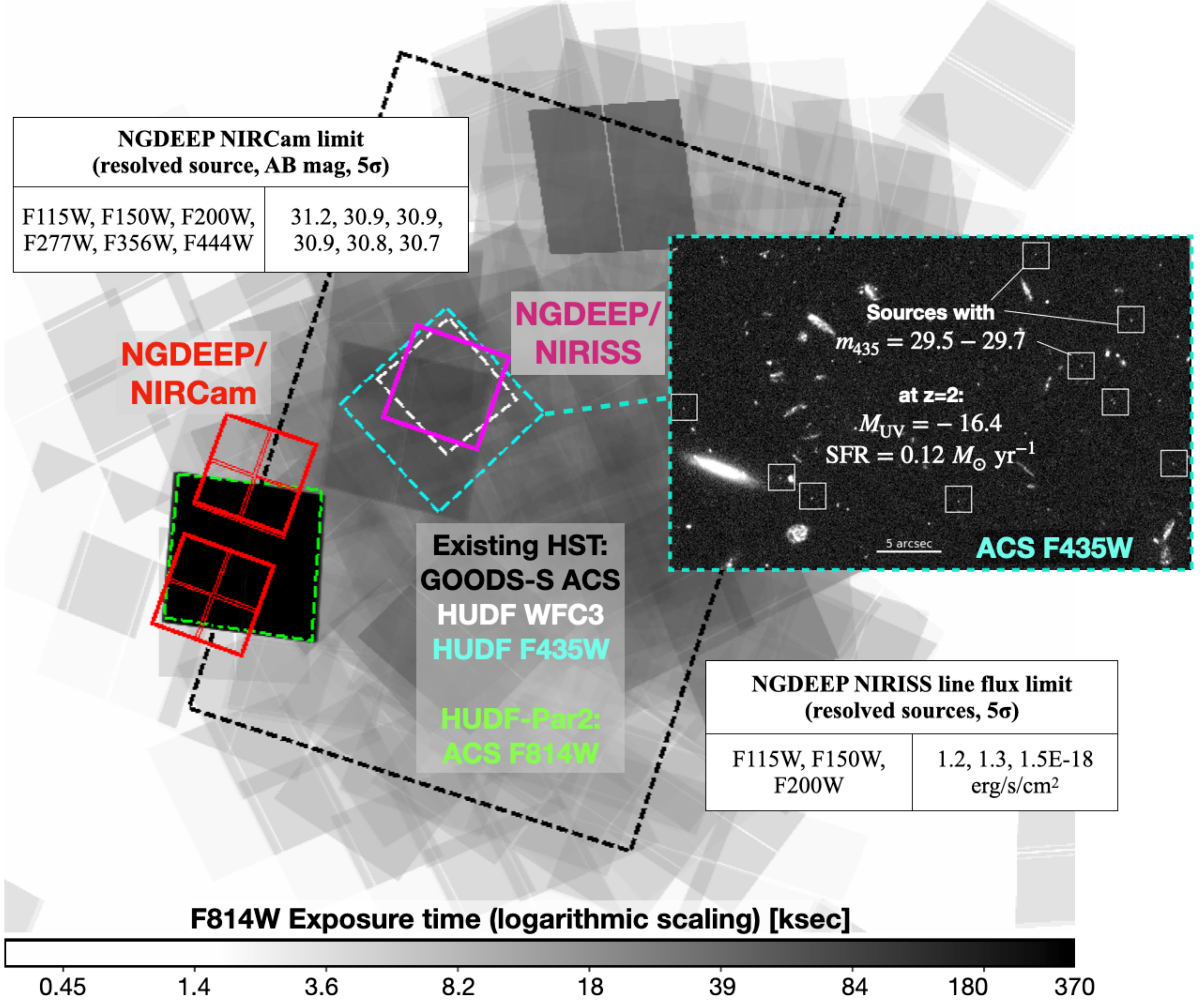}
\caption{The layout of NGDEEP.  The background grayscale image is the exposure time map of {\it HST}/ACS F814W imaging, shown with logarithmic scaling as given by the scale bar.  The NGDEEP NIRISS (NGDEEP-NIS) pointing is shown by the magenta region (in the HUDF), while NGDEEP NIRCam (NGDEEP-NRC) is shown by the red regions (in the HUDF Par2 parallel field) for one of our two position angles. Extremely deep {\it HST}/ACS F435W imaging exists in the HUDF \citep{illingworth13}, achieving depths of 29.6 AB mag (5$\sigma$, \citealt{illingworth13}), while the deepest {\it HST}/ACS F814W imaging anywhere in the sky is in the HUDF Par2 field.  The inset tables show our expected line-flux sensitivities for NGDEEP-NIS, and continuum sensitivities for NGDEEP-NRC.  The image inset shows a portion of the ACS F435W imaging, with sources near the detection limit indicated by white boxes.  At $z= 2$ the $m_{435}$ magnitudes correspond to rest-frame UV magnitudes of $M_\mathrm{UV} = -16.7$, or an unattenuated SFR of $0.12$~$M_\odot$ yr$^{-1}$ (\citealt{kennicutt12}).  This SFR would produce H$\alpha$ fluxes detectable by our NIRISS spectroscopy. }
\label{fig:field_layout}
\end{figure*}

\subsection{Field Choice and Orientation}\label{sec:pas}
We center the NGDEEP-NIS observations on the HUDF to fully exploit the large investment of {\it HST} observations. Imaging with \HST/ACS $B$-band \citep[F435W;][]{illingworth13} reaches a limiting depth of m$_{5\sigma\mathrm{,F435W}}=29.6$, sufficient to detect galaxies with UV-based SFRs to $\sim$0.1 M\sol\ yr$^{-1}$ at $z \sim$ 2, which matches the NIRISS H$\alpha$ SFR limit of the NGDEEP-NIS observations. Extremely deep WFC3/IR imaging in the same region provides critical constraints on the stellar masses of the star-forming galaxies (to log [M$_*$/M\sol] $\sim$ 7.3 at $z =$ 2).  
These data in the HUDF are the deepest available.
The separation between NIRISS and NIRCam on the {\it JWST} field-of-view is similar to that of WFC3 and ACS, placing our coordinated parallels on the HUDF05-02 parallel field (HUDF-Par2). This field has been observed repeatedly with ACS while NICMOS and WFC3 observed the HUDF \citep{stiavelli05,illingworth09}, building the deepest F814W imaging on the sky ($m_{5\sigma,\mathrm{F814W}}=30.0$).  While NIRCam alone can select $z >$ 9 galaxies, these candidates will be strengthened by non-detections in this ultra-deep optical image, which covers $\gtrsim$50\% of our deepest NIRCam data, and the deep F814W data also allows dropout-selection to $z \sim$ 7--9.

Centering the NIRISS primary data on the HUDF and placing NIRCam on HUDF-Par2 results in an observatory V3 position angle V3\_PA = $\sim$70$^{\circ}$.  We obtain the NIRISS WFSS with both the row (R) and column (C) grisms to provide two-orient observations.  However based on our trade studies, simulated spectroscopy, and previous work in the literature \citep{pirzkal04,ryan18}, an additional position angle separated by $\sim\!3^{\circ}$ will significantly improve the contamination modeling, emission-line identification, and map-reconstruction without sacrificing the spatial coverage of the HUDF. However, splitting the observations across two position angles results in less area covered at total depth for NIRCam. We therefore limit the difference in position angle to $\Delta$(V3\_PA)=3$^{\circ}$, the minimum acceptable for NIRISS without sacrificing full-depth coverage for NIRCam. Therefore, each source in the HUDF will be observed in four distinct position angles (V3\_PA$=$67,70$^{\circ}$, R and C grism). The deepest portion of the NIRCam parallel will cover 5 arcmin$^2$ and will reach $\sim$0.5 mag shallower over 10 arcmin$^2$.

In Figure~\ref{fig:field_layout}, we present a visualization of the NGDEEP-NIS and NGDEEP-NRC field centroids and orientations. The figure also provides a summary of some of the existing \HST\ datasets overlapping the NGDEEP observations. 

\subsection{NGDEEP Epochs 1 and 2}\label{sec:program_update}
The strict position angle requirements for NGDEEP result in a limited window of $\sim$10--16 days during which the program can be scheduled. Unfortunately, on January 20, 2023, NIRISS experienced an unexpected software timeout. All primary and coordinated parallel NIRISS observations were temporarily suspended while the observatory and instrument teams diagnosed the problem. When NIRISS observations resumed ten days later, there was not enough time to observe the entire program before the NGDEEP observing window closed on February 6.
Instead, all observations from a single position angle (V3\_PA=70$^{\circ}$) were obtained in the remaining time. This Epoch 1 of NGDEEP includes all grisms and filters but half of the total integration time for both instruments. In Sections~\ref{sec:nis_obs} and \ref{sec:nrc_obs}, we present the sensitivity and limiting magnitude estimates that correspond to the total program time. Epoch 2 (V3\_PA=67$^{\circ}$) will be observed in January 2024.

\begin{deluxetable}{cccc}
\tablecaption{NGDEEP Line Sensitivities and Limiting Magnitudes}\label{tab:exptimes}
\tablehead{
\colhead{Filter} & \colhead{Specification\tablenotemark{a}} & \colhead{Exposure Time (h)} & \colhead{Sensitivity} } 
\startdata
\multicolumn{3}{l}{NGDEEP-NIS} & (erg s$^{-1}$ cm$^{-2}$)\\
F115W & NIS/20/6 & 52.2 & $1.2\times 10^{-18}$ \\
F150W & NIS/20/8 & 34.8 & $1.3\times 10^{-18}$ \\
F200W & NIS/20/6 & 17.4 & $1.5\times 10^{-18}$ \\
\hline
\multicolumn{3}{l}{NGDEEP-NRC} & (AB mag) \\
F115W & DEEP8/4/7 & 53.9 & 31.2 \\
F150W & DEEP8/5/7 & 23.0 & 30.9 \\
F200W & DEEP8/4/7 & 18.0 & 30.9 \\
F277W & DEEP8/5/7 & 22.3 & 30.9 \\
F356W & DEEP8/4/7 & 20.8 & 30.8 \\
F444W & DEEP8/4/7 & 51.8 & 30.7 \\
\enddata
\tablenotetext{a}{The observing specification of the deepest observations in each filter/grism, listed as Readout pattern/groups per integration/integrations per exposure. The NGDEEP-NRC F115W, F150W, F200W and F356W filters have additional imaging, observed in parallel to the NIRISS-NIS direct imaging and taken with the SHALLOW4 readout pattern, 3 groups and 1 integration.}
\tablecomments{Estimates for NGDEEP-NIS $5\sigma$ integrated emission line sensitivities and NGDEEP-NRC $5\sigma$ resolved-source limiting magnitudes are from v2.0 of the \JWST\ ETC. These estimates are based on the full survey depth, half of which (Epoch 1) was observed in February 2023.}
\end{deluxetable}

\begin{figure*}
\centering
\gridline{\fig{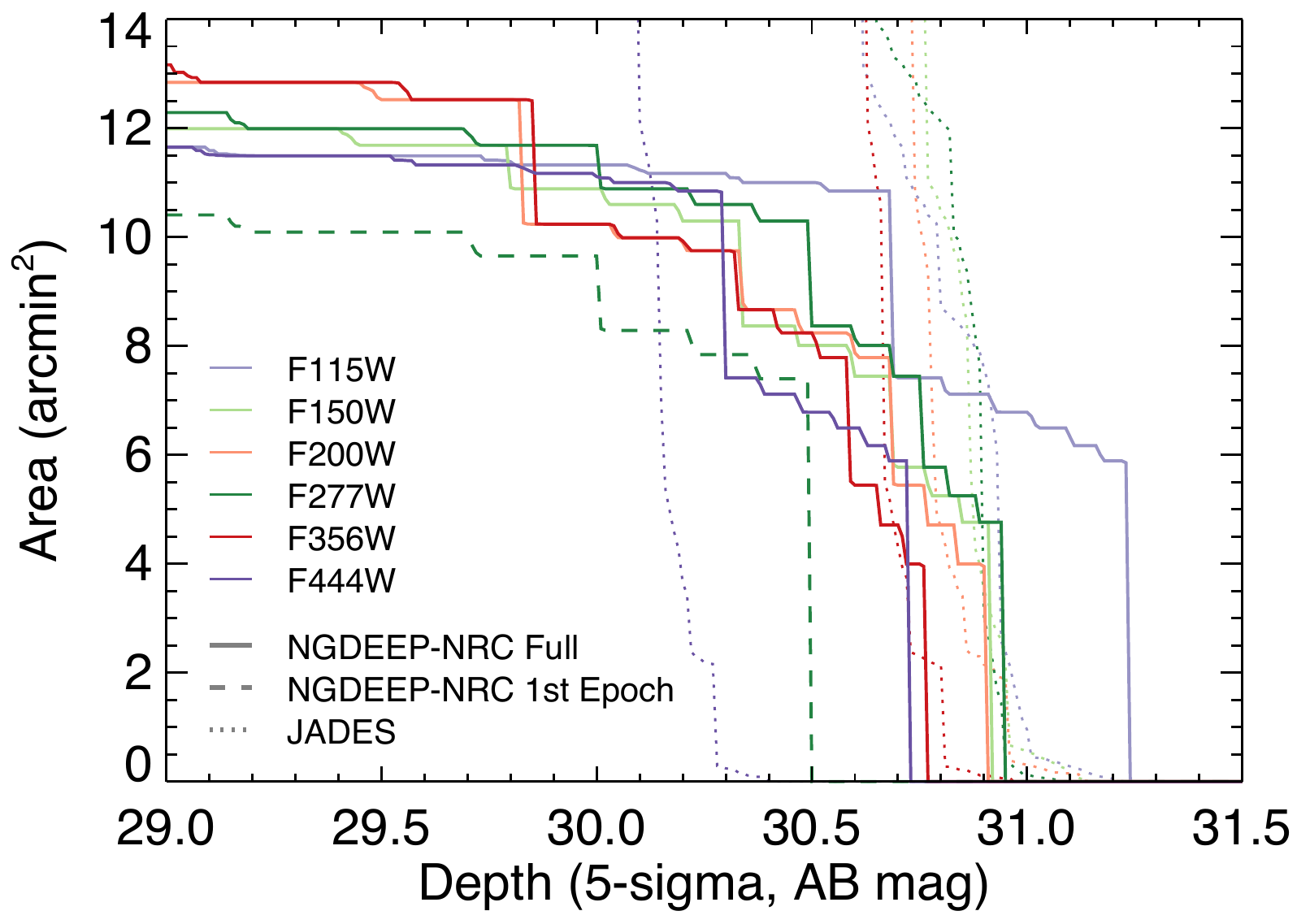}{0.5\textwidth}{}
          \fig{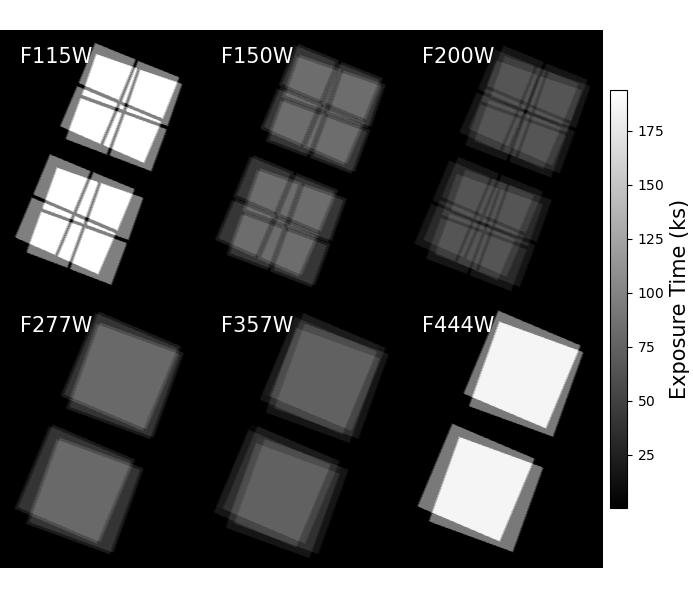}{0.43\textwidth}{}}
\vspace{-9mm}
\caption{
\textit{Left:} 
The area covered in each NGDEEP-NRC filter as a function of $5\sigma$  depth measured for resolved sources. The solid lines denote the full survey (Epochs 1 and 2 covering both position angles), and the dashed line shows the depth and area coverage achieved in Epoch 1 (V3\_PA=70$^{\circ}$) for F277W as an example. The dotted lines show the depths that will be achieved by the JADES deep field observations in the HUDF once the survey is completed. While JADES covers a larger area, NGDEEP-NRC will go $\sim$0.5 mag deeper in F444W and $\sim$0.3 mag deeper in F115W, allowing for the detection of $z>9.5$ \lya-breaks in galaxies at the detection limit in the F150W, F200W and F277W. As a treasury program, NGDEEP will also be public immediately, providing the community the opportunity to explore a \JWST\ deep field right away in Cycle 1.
\textit{Right:} Exposure time maps in $10^3$ seconds for each of the six NGDEEP-NRC filters. The overlapping position angles, NIRISS filter FWHM-dependent direct imaging dithers, and short wavelength detector gaps result in the multiple depth tiers that are present in the left panel.
\label{fig:nirtech}}
\end{figure*}

\subsection{NGDEEP-NIS Observations}\label{sec:nis_obs}
The NGDEEP observations obtained in the HUDF with NIRISS (NGDEEP-NIS) will provide deep $R\sim150$ spectroscopy over $1-2.2$\micron. The spectra will detect emission lines from $>$1000 galaxies, many with multiple lines, down to a limiting emission line flux of $1.2-1.5\times10^{-18}$ erg s$^{-1}$ cm$^{-2}$ (5$\sigma$). This is not only much deeper than previous {\it HST} grism surveys \citep[e.g.][]{pirzkal17,momcheva16}, but extends the upper end of the wavelength range covered from 1.7 $\mu$m to 2.2 $\mu$m (Figure~\ref{fig:nis_sims}). 


Multi-line diagnostics for metallicity and dust content requires sampling [\ion{O}{2}], [\ion{O}{3}], H$\beta$, and H$\alpha$. NGDEEP-NIS will obtain F115W, F150W, and F200W spectroscopy in order to examine the redshift evolution of these metrics (for $1.6 < z < 2.5$). As mentioned above, we will obtain spectroscopy in both the R and C grisms as well as at two PAs separated by 3$^{\circ}$ (V3\_PA$=$67,70$^{\circ}$). The grism exposures will be obtained with the NIS readout pattern, 20 groups per integration and 6 integrations per exposures (8 integrations for F150W). We will use a 3 point dither pattern optimized for NIRISS-NIRCam coordinated parallel observing. 
The total exposure times in these three filters will be 190, 86, and 63 ks, respectively. Based on 1D outputs from the \JWST\ Exposure Time Calculator (ETC, v2.0) and optimal extractions of our simulated NGDEEP-NIS observations (see \S\ref{sec:nis_sims}), NGDEEP-NIS will reach $1.2-1.5 \times 10^{-18}$ erg s$^{-1}$ cm$^{-2}$\ ($5\sigma$). This allows us to measure metallicities at $\sim$0.2 dex precision down to log($M_*$/M\sol)$=$7 at $z=$ 2 (see \S\ref{sec:nis_science}).  We summarize the NGDEEP-NIS observing specifications, exposure times, and line sensitivities in Table~\ref{tab:exptimes}. These observations include direct imaging in the F115W, F150W, and F200W filters with total exposure times of 10.8, 3.6, and 3.6 ks, reaching m$_{AB}=$ 29.5,  29.0, and 29.1 (3 $\sigma$), respectively, deep enough to identify the faint continuum of contaminating sources.
Our NGDEEP-NIS simulations (\S 5) demonstrate we will detect emission lines ($\geq$ 3$\sigma$) in $>$1000 galaxies (see Table~\ref{tab:ncounts}).

While the NIRSpec GTO team targets this field with NIRSpec (PI: N Luetzgendorf, PID 1210), examination of their pre-launch APT file finds that only $\sim$80 objects in the HUDF will receive slits with $t\!=$8 ks. Early results from these GTO observations are already revealing the exquisite sensitivity of \JWST\ spectroscopy, with spectroscopic confirmations of four galaxies at $z>10$ \citep{curtis-lake2022,robertson2022b}. NGDEEP will expand this legacy, with $>$1000 galaxies receiving 60--190 ks spectroscopic integrations.

\subsection{NGDEEP-NRC Observations}\label{sec:nrc_obs}

The NGDEEP observations obtained in the HUDF-Par2 field with NIRCam (NGDEEP-NRC) will  provide deep imaging from $\sim$1--5 $\micron$. Imaging in F115W$+$F150W$+$F200W will sample below the Ly$\alpha$ break and the rest-UV continuum for galaxies at $z>9$. We simultaneously observe with F277W$+$F356W$+$F444W to fully sample the spectral energy distribution (SED) and minimize sample contamination. The NIRCam observations will enable the discovery of $\sim$30--100 galaxies at $z \gtrsim$ 11 and constrain the faint-end of the rest-UV luminosity function where stellar feedback model predictions differ the most. NGDEEP-NRC will also constrain black hole formation and probe the morphological transformation of galaxies. We discuss the main NGDEEP-NRC science goals in \S\ref{sec:nrc_science}.

The NIRCam strategy is set by the primary NIRISS observations, though we distribute the integration time to achieve approximately uniform sensitivity (30.7--30.9 mag; 5$\sigma$) in all filters. We increase the time in F115W (and F444W simultaneously) to detect $\sim$0.3--0.4 mag Ly$\alpha$-breaks at our limit of $m\!=$30.7--30.9. These data will allow for a robust selection of galaxies at $9<z<13$ that are detected in $\geq$4 filters ($>$2 mag deeper than CEERS and up to 0.1--0.5 mag deeper than the JADES GTO Program). 
The individual exposure times are $\sim$5.1 ks (F115W, F200W, F356W, and F444W) and $\sim$6.7 ks (F150W, F277W) with the DEEP8 readout, 4 groups per integration (5 for F150W, F277W), and 7 integrations per exposure.  Images taken in parallel to NIRISS direct imaging have t$=$150s using SHALLOW4 readout to allow for 3 groups. The final exposure times will be 60--84 ks (F150W, F200W, F277W, and F366W) and 170-180 ks (F115W and F444W). We estimate depths with the ETC, assuming expected resolved sizes (FWHM $\sim$0.07\arcs\ at $m=30.6$; \citealt{kawamata18}). 

The combination of imaging at two position angles separated by 3$^{\circ}$ will result in tiers of depth in the NGDEEP-NRC mosaic. We present the anticipated area as a function of depth in the left panel of Figure~\ref{fig:nirtech}, where these depth tiers are evident as a set of step functions for the NGDEEP observations.
All six bands are observed in the deepest tier to at least 30.7 mag. The filters F150W, F200W, and F277W will reach 30.9 mag, and F115W will achieve a depth of 31.2 mag.  At an area of 5 arcmin$^2$, NGDEEP-NRC is expected to be comparable in depth to the deepest tier of JADES in F150W, F200W, F277W, and F356W.  However, NGDEEP-NRC is $\sim$0.3 mag deeper than JADES\footnote{NGDEEP vs.\ GTO Programs: 
We derived JADES and MDS exposure times and depths directly from their APT files, assuming 5$\sigma$ resolved sources for comparison with NGDEEP-NRC depths. Our estimates of the depths differ in places from the summaries in \citet[][JADES]{williams2018}  and \citet[][MDS]{perez-gonzalez23}.  
} in the important F115W $z \sim$ 10 dropout filter  (194 ks over our deepest 5 arcmin$^2$ compared with 143 ks for JADES), allowing robust candidate galaxy selection to the limiting magnitude in the redder filters.  NGDEEP-NRC will also be $\sim$0.5 mag deeper in F444W (186 ks compared with 94 ks), a crucial filter for limiting contamination by low-redshift objects.  JADES is shown as the dotted lines in Figure~\ref{fig:nirtech}.

\begin{figure*}
\epsscale{1.1}
\plotone{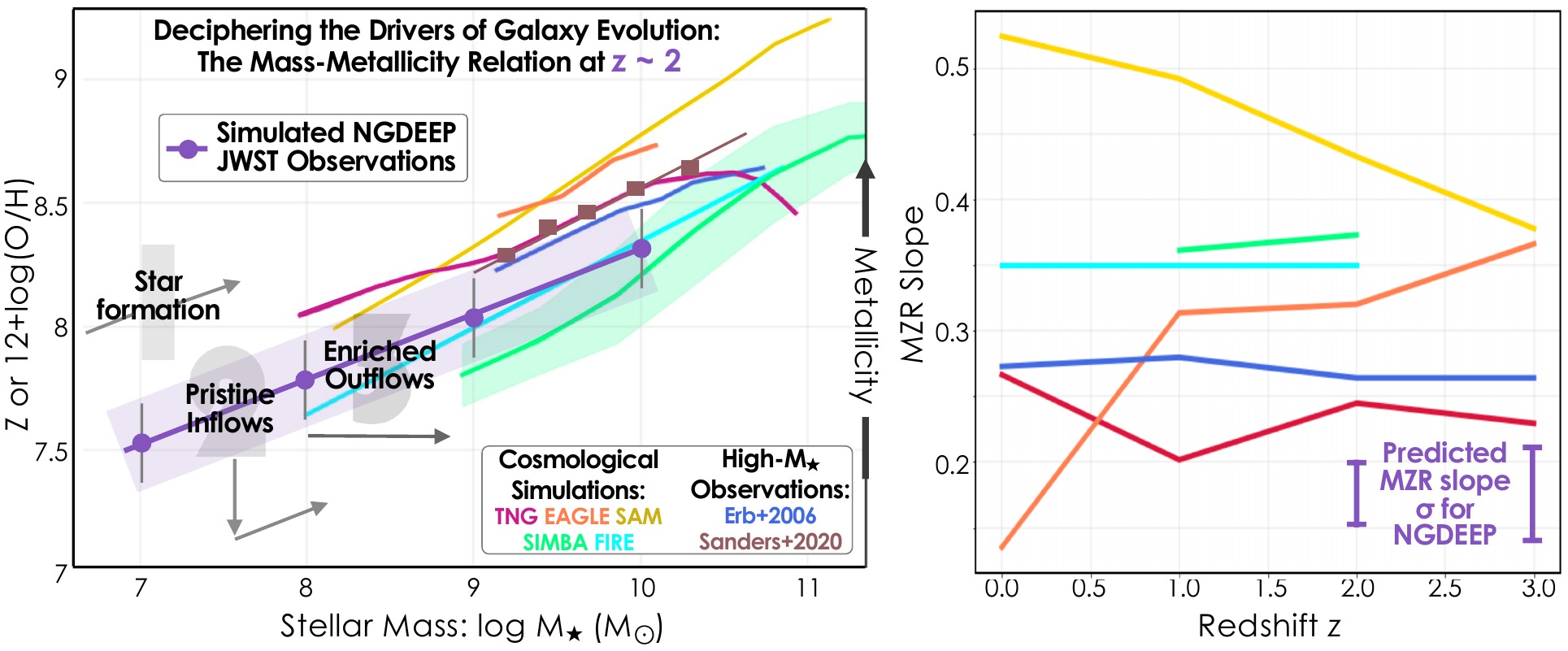}
\caption{Understanding the details of the MZR is fundamental for 
theories of galaxy evolution.  \textit{Left}:  arrows show that (1) metal abundance increases with the nucleosynthetic yields of star formation, (2) accretion of metal-poor gas both dilutes metals and can trigger star formation, 
(3) outflows remove enriched gas with an efficiency that is inversely proportional to the 
galaxy gravitational well.   The purple line and swath show the MZR and its scatter measured from synthetic NIRISS spectra, derived from our simulations:  NGDEEP will significantly extend the MZR mass coverage by $\sim$2 dex. 
\textit{Right}: The diversity in the evolution of the MZR slope predicted by simulations with different assumptions about feedback (with color scheme as in the left panel): NGDEEP will measure this slope and scatter, and thereby constrain the physics of feedback.}
\label{fig:MZR}
\end{figure*}

The MIRI Deep Survey (MDS) operated by the MIRI European Consortium Team 
is also obtaining deep observations with MIRI on the HUDF. Their parallel NIRCam imaging includes 55 ks of integration for F115W, F150W, F277W, and F356W. We note that three of the four planned visits were observed in December 2022, with the final visit scheduled for December 2023. Additionally, one of the three visits was observed at a slightly different position angle. As a result, the current set of MDS NIRCam observations include 55 ks in F115W+F277W and 28 ks in F150W+F356W, reaching depths of m$\sim$30.0-30.2 in all filters and m$\sim$30.7 in F277W. There is an area of $\sim$8 arcmin$^2$ in common between the two filter sets, $\sim$3.7 arcmin$^2$ of which overlaps with the NGDEEP-NRC footprint. Early results from \citet{perez-gonzalez23} have identified 45 new, faint $z>8$ candidates, highlighting the potential offered by the increased depth and filter coverage that NGDEEP-NRC brings to this region.   

Based on the predictions of a range of models, NGDEEP-NRC will discover $\sim$50--100$+$ galaxies at $z \sim 10$, and up to 60, 25, 12 and 10 galaxies at $z \sim$ 11, 12, 13 and 14--15, respectively (see \S\ref{sec:nrc_science}), sufficient to distinguish between competing models for high-$z$ galaxy evolution.
In \S\ref{sec:nrc_sims}, we also demonstrate that NGDEEP-NRC's $\sim$2 mag increase in depth compared to CEERS is essential to detect galaxies at $z=10-13$.



\section{NGDEEP-NIS Science Cases: Star-Formation, Enrichment, and Feedback at $\lowercase{z} =$ 0.5--4}\label{sec:nis_science}

\subsection{Constraining Feedback via the Low--Mass end of the Mass--Metallicity Relation} \label{sec:mzr}

A fundamental probe of galaxy evolution is the stellar-mass versus gas-phase metallicity relation \citep[MZR; e.g.,][]{tremonti04,maiolino08}.  
The MZR is thought to result from the general growth of galaxies over time,
where they increase their both their stellar mass and abundance of metals produced, but are also sensitive to competing feedback processes.   
Figure~\ref{fig:MZR} shows the expected tight, positive MZR for NGDEEP-NIS.
The MZR slope, normalization, and intrinsic scatter is shaped by galaxy feedback 
processes, where metals are created by star formation, ejected by outflows, 
and diluted by gas infall  \citep{tremonti04,zahid14}.  
Further, the MZR is observed to exist out to at least $z{\sim}3$--4 
\citep[e.g.,][]{maiolino19,sanders20}, where it has been observed to have a lower normalization than at $z\sim0$.
This redshift evolution is thought to result from a number of effects including stellar and 
AGN feedback-driven outflows, changing stellar yields, and star formation histories.
Simulations predict the MZR in detail \citep[e.g.,][]{dave11,torrey14}, but show significant tension.  
As depicted in the right hand panel of Figure~\ref{fig:MZR}, different state-of-the-art models 
predict staggeringly different redshift evolution of the MZR slope at low-masses:
predictions vary by more than 0.3 dex in $d\log Z/d\log M_\ast$.  
Progress requires robust metallicity measurements in low-mass galaxies across a range of 
redshifts. 

\begin{figure*}
\epsscale{1.15}
\plottwo{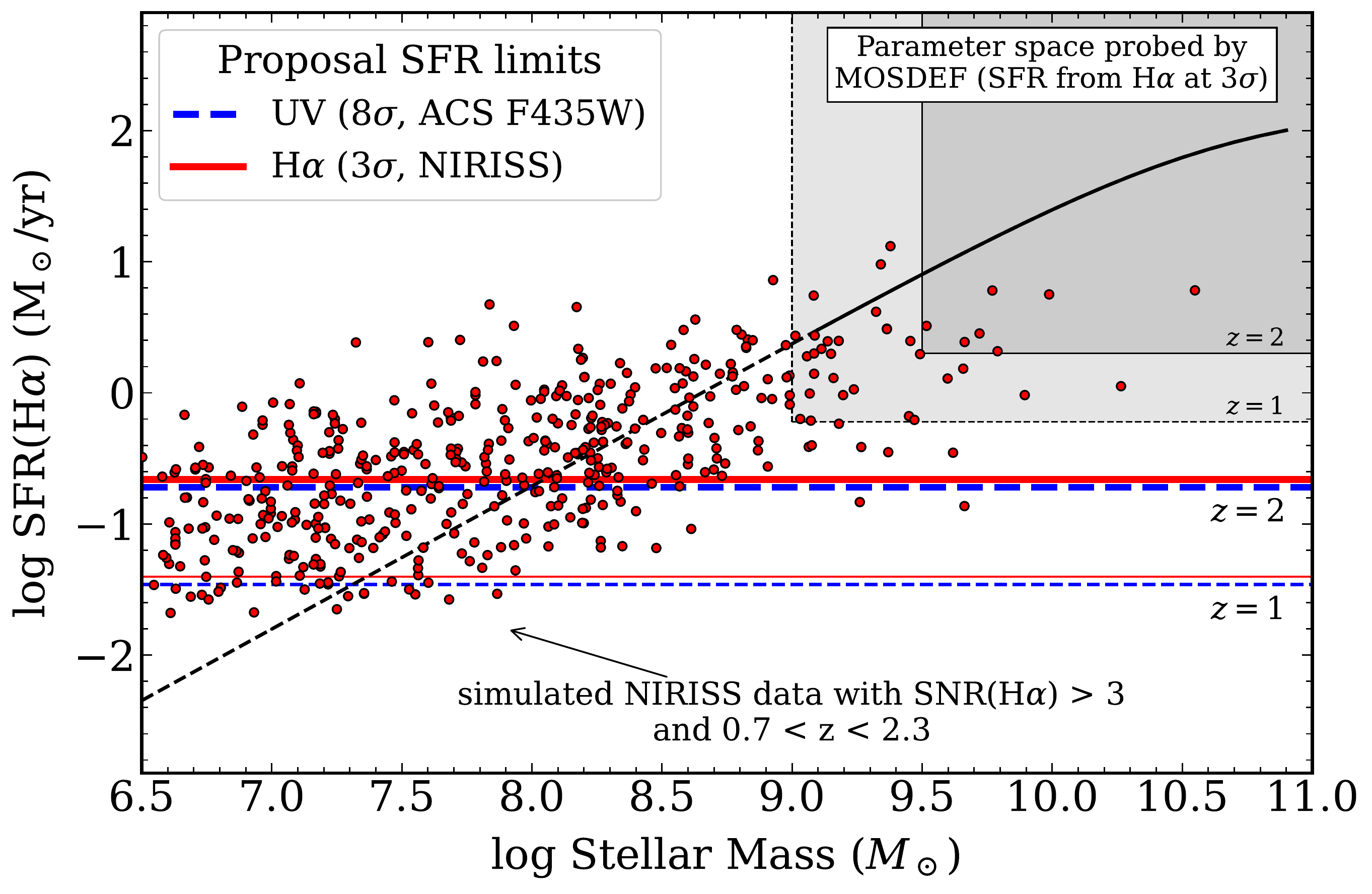}{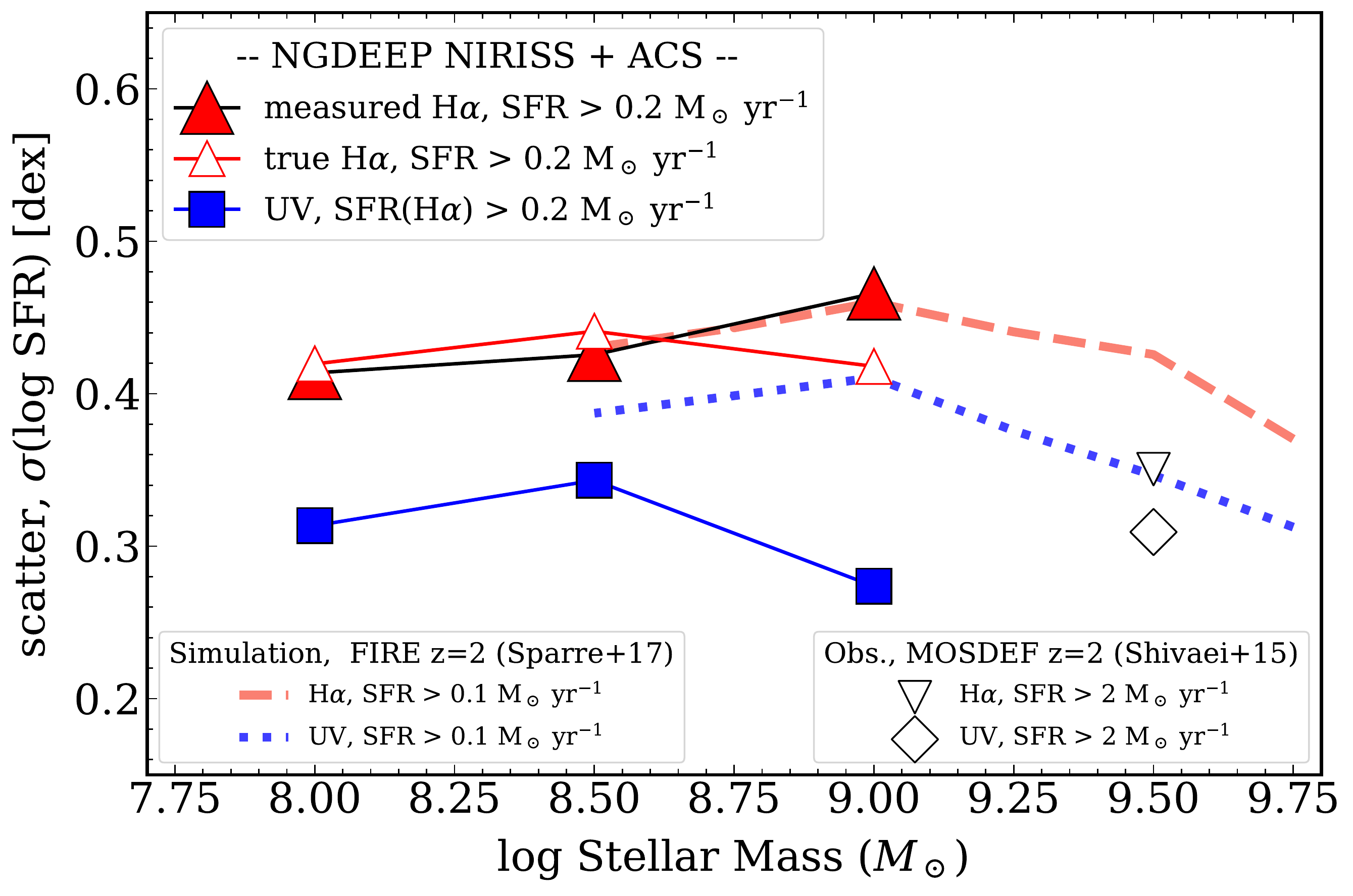}
\caption{NGDEEP-NIS will enable the measurement of the SFR-Stellar Mass (SFMS) relation and its scatter at $z=1-2$, for galaxies 10$-$100$\times$ fainter than previous surveys (e.g., MOSDEF).
\textit{Left:} The predicted SFR-Stellar Mass (SFMS) relation for NGDEEP-NIS (small red circles). NIRISS will derive SFRs from \Ha\ for galaxies to SFR$=0.22$ (0.040) M\sol\ yr$^{-1}$ at $z=2$ ($z=1$). The \Ha-derived SFR limits are well matched to the UV-derived SFRs from the HUDF ACS F435W imaging. The black-solid line shows the SFMS at $z=2$ \citep{tomczak14}, the black-dashed line is the extrapolation to log$M_{*}/M_{\odot} < 9$.
\textit{Right:} The NGDEEP-NIS \Ha-/UV-SFRs will measure the scatter in the SFMS. The data points show the scatter from the SFMS for galaxies above a given stellar mass. For NIRISS, we show the true scatter (red open triangles) compared to the measured scatter (large red-filled triangles), which include measurement errors. The difference in the scatter between UV and \Ha\ SFRs provides a strong constraint on feedback in simulations: this is predicted to be strongest at the lower stellar masses (log$M_{*}/M_{\odot} < 9$, see \citealt{sparre17}) probed by NGDEEP-NIS.
\label{fig:sfms}}
\end{figure*}

\begin{deluxetable}{cccc}
\tablecaption{Expected Emission Line Counts in NGDEEP-NIS}\label{tab:ncounts}
\tablehead{
\colhead{Line or Index} & \colhead{Number} & \colhead{Number} & \colhead{Redshift Range}  \\
\colhead{} & \colhead{($3\sigma$)} & \colhead{($5\sigma$)} & \colhead{}
}
\startdata
Any line & 2266 & 1601\\
Ly$\alpha$ \W1216   & 6     & 1 & 7.22--17.09 \\
\MgII\ \W\W2796,2804& 40    & 19 & 2.57--6.87 \\
\OII\ \W\W3727,3730 & 464   & 307 & 1.68--4.90 \\
\NeIII\ \W3870      & 40    & 8 & 1.58--4.68 \\
\Hb\ \W4863         & 289   & 162 & 1.06--3.52 \\
\OIII\ \W5008       & 837   & 618 & 1.00--3.39 \\ 
\Ha\  \W6565        & 965   & 676 & 0.52--2.35 \\
\SII\ \W\W6718,6732 & 156   & 96 & 0.49--2.27 \\
\hline
\hline
\multicolumn{3}{c}{Number of Galaxies with $\Delta z < 0.2$ dex} \\
\hline
\multicolumn{2}{c}{$z > 1$} & 361 \\
\multicolumn{2}{c}{$z > 2$} & 275 \\
\multicolumn{2}{c}{$z > 3$} & 98 \\
\enddata
\tablecomments{Number of emission lines (or robust multi-line metallicities) in our end-to-end NGDEEP-NIS simulations at 3 and 5$\sigma$ significance, using the JADES galaxy mock catalog as input (see \S\ref{sec:nis_sims}). We note that the \lya\ predictions debend on the assumed IGM attenuation and may be underestimates, as several \lya\ detections at $z>7$ have already been confirmed
\citep[e.g.,][]{vanzella11,castellano18,tilvi20,jung20,jung22,larson22}.
Note that wavelengths are in rest-frame vacuum. Doublets will be blended.}
\end{deluxetable}

NGDEEP-NIS will obtain NIRISS slitless spectroscopy covering an observed range of 1.0--2.2~\micron\ 
with the sensitivity to detect emission lines as faint as $\sim 10^{-18}$ erg s$^{-1}$ cm$^{-2}$.
This will enable the measurement of galaxy gas-phase metallicities using well-calibrated 
rest-frame optical emission-line diagnostics from 
$R_{23} \equiv$~(\OII\W3727 + \OIII\W5008)/\Hb\ at $1.7<z<3.4$ and 
Ne3O2 $\equiv$~\NeIII\W3868$ / $\OII\W3727 to $z\lesssim5$ \citep[e.g.,][]{maiolino19}.  
While most MZR studies only probe the most massive galaxies in targeted surveys, NGDEEP-NIS will 
measure metallicities for an unbiased galaxy sample (i.e., no preselection) with stellar masses of $\log M_\ast/M_\odot{\gtrsim}7$ at these redshifts. 
Recently, \cite{Li22} used stacks of \textit{JWST} observations of 55 $z\sim2-3$ galaxies to reveal our first glimpse of the evolution of the MZR out to cosmic noon.
This work hints at a shallower slope for low-mass galaxies ($M_\star < 10^9 M_\odot$),
possibly due to the dominance of different feedback processes in the low-mass regime.
NGDEEP-NIS will significantly increase the sample size of low-mass galaxies on the MZR
at cosmic noon, allowing us to discern any turnover resulting from varying feedback sources.

Figure~\ref{fig:MZR} shows the MZR (slope, normalization, and scatter) from our NGDEEP-NIS simulations 
using $R_{23}$, showing constraints to $\lesssim$0.2 dex across our mass range.  
Comparing to simulation predictions, the NGDEEP-NIS MZR will place stringent empirical constraints on 
the appropriate feedback prescriptions that should be incorporated into models. 
Joint constraints on the low-mass end of the stellar mass function and the MZR can constrain the 
mass-loading factors of stellar- and AGN-driven winds, and test if feedback is ejective or preventative, thereby constraining the energy content of winds. 
NGDEEP-NIS measurements of the MZR will thus provide critical benchmarks for the next generation of galaxy formation simulations in and after the {\it JWST} era.

\subsection{Constraints on Bursty Star Formation from rest-UV and  H$\alpha$ measurements}\label{sec:burstySF}

A powerful capability of NGDEEP-NIS is that it combines NIRISS measurements of the H$\alpha$ recombination line for galaxies at $0.7<z<2.3$ with rest-UV measurements from HUDF ACS F435W (the deepest $B$-band imaging anywhere on the sky, $m_{5\sigma}=29.6$; \citealt{illingworth13}).  Figure~\ref{fig:sfms} shows that the NIRISS H$\alpha$ line flux limit is well matched to the ACS rest-UV limit:  both will detect SFRs from galaxies down to 0.22 (0.04) $M_\odot$ yr$^{-1}$ at $z=2$ ($z=1$).  With NGDEEP-NIS we will make a first measurement of the stellar-mass--SFR relation down to $\log M_\ast/M_\odot=7$  at $z=1$--2. 

\begin{figure*}
\plotone{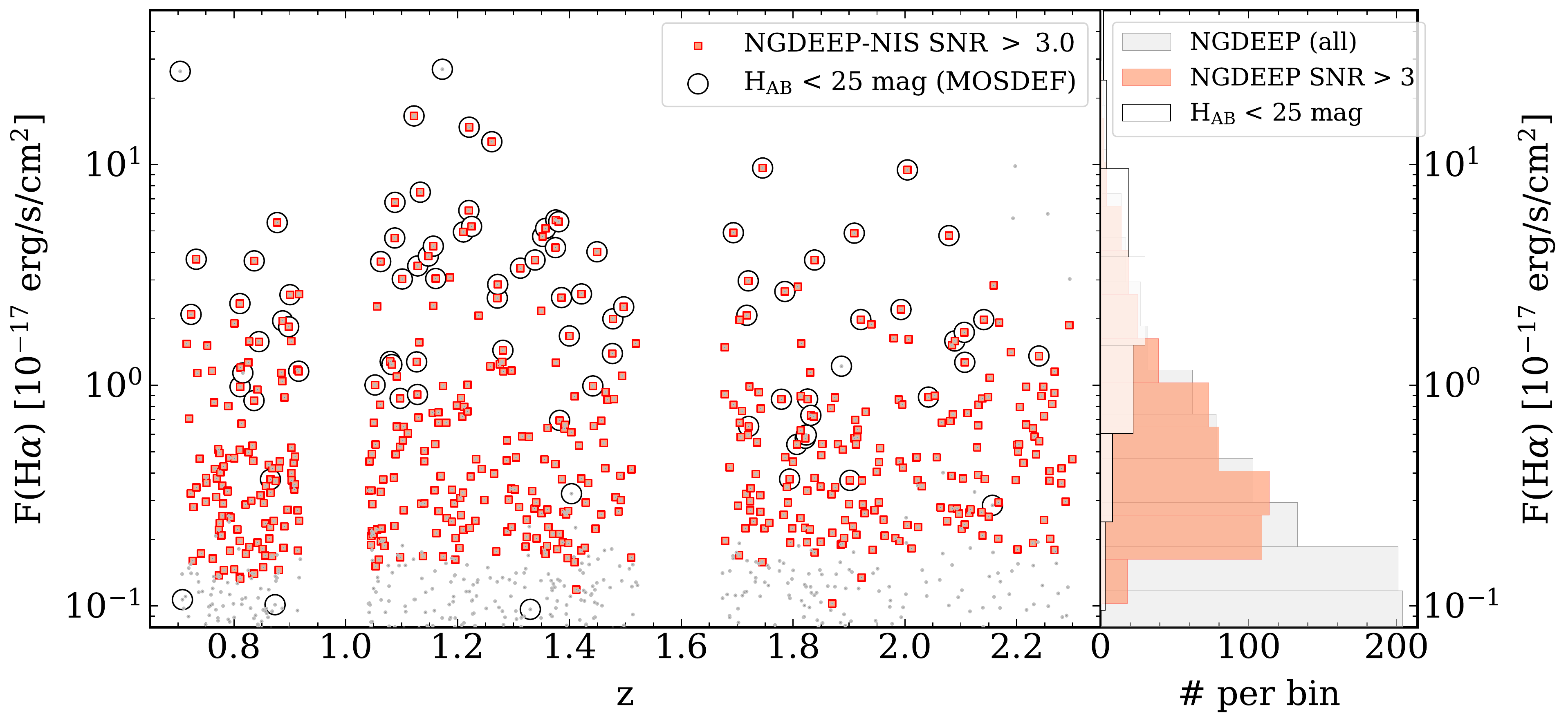}
\caption{ 
Redshift versus \Ha\ flux for galaxies in simulated NGDEEP-NIS data. The small grey points represent all sources in the simulation. The red squares are simulated objects with NIRISS-detected \Ha\ emission (SNR~$> 3\sigma$). NGDEEP will measure faint \Ha\ emission from $>$900 sources from $0.7 < z < 2.3$ (see Table~\ref{tab:ncounts}). The open circles are sources that would have been targeted by MOSDEF, a dedicated survey using the 10m Keck telescope. Surveys like MOSDEF must preselect objects and are subject to slit crowding: MOSDEF only targeted galaxies with $H_{160} < 25$ AB mag and had a flux limit of $F \simeq 10^{-17}$ erg s$^{-1}$ cm$^{-2}$ ($3\sigma$). MOSDEF would be sensitive to fewer than 70 of the NGDEEP-NIS sources (at higher stellar masses). 
\label{fig:flux_ha}}
\end{figure*}

These NGDEEP SFR observations will provide diagnostics on the stochasticity of star formation, another strong probe of feedback. The amplitude and characteristic timescale of star formation variability is a test of state-of-the-art cosmological hydro simulations which implement star formation and feedback differently \citep{Iyer:2020}, where increased star formation efficiency and stronger feedback produce more stochastic star formation \citep{hopkins:2018}.   
The H$\alpha$ and UV emission are sensitive to different star-formation timescales:  the UV continuum is emitted by massive (OB) stars, and probes SF variations on $\sim$100~Myr timescales; H$\alpha$ emission requires ionization from the most massive (O) stars, and probes 5-10~Myr timescales \citep{kennicutt12}.   Observations of H$\alpha$ and UV SFRs are therefore sensitive to stochasticity and bursts, the effects of which are strongest in lower mass galaxies \citep{weisz12,Broussard:2019}.  

With NGDEEP-NIS we will combine the H$\alpha$ emission from NIRISS and rest-UV from ACS to study star-formation variability in low-mass galaxies ($\log M_\ast/M_\odot{\simeq}7$) at $0.7{<}z{<}2.3$.  NGDEEP-NIS enables the 
measurement of the scatter in these SFRs for galaxies above a fixed SFR and stellar mass, achieving results for galaxies $10\times$ fainter than previous work (MOSDEF, see Figure~\ref{fig:flux_ha}, \citealt{shivaei15}), where theory predicts the SFR scatter is more pronounced \citep{sparre17,Iyer:2020}. The right panel of Figure~\ref{fig:sfms} shows we are able to recover the intrinsic SFR scatter $\sigma$=0.45 for H$\alpha$ and 0.3--0.35 for the UV for our simulated dataset.   If the real difference between the H$\alpha$ and UV SFR scatter is larger than predicted by models, then it would imply stronger ejective feedback is needed to regulate star formation in low-mass systems.    NGDEEP-NIS is the only survey sensitive enough to measure SFRs to these limits in both H-recombination lines and UV continua for galaxies at $z{\sim}1$--2.

\subsection{Physical conditions of galaxies in the HUDF}\label{sec:physconditions}
The unbiased nature of the NIRISS slitless spectroscopy means the data probe every galaxy in the HUDF.  This will enable additional science for galaxies beyond that articulated here. This includes galaxies with additional diagnostic lines (see Table~\ref{tab:ncounts}), such as rest-UV spectra of faint galaxies at $z{>}4$.   Therefore the dataset here has a legacy value with which to study galaxies in this {\it JWST}$+${\it HST} deep field.

\begin{figure*}
\plotone{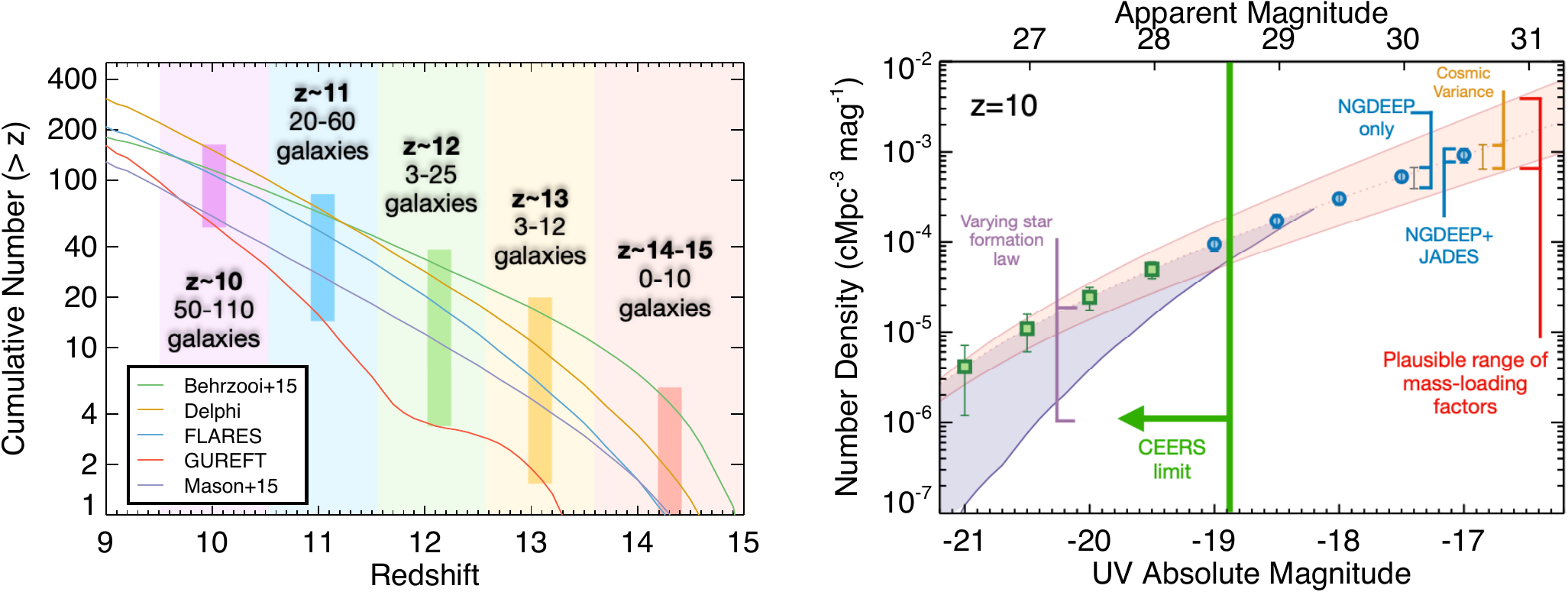}
\caption{\textit{Left:} Predicted cumulative number of high-redshift galaxies detected across the NGDEEP NIRCam imaging mosaic. Simulations \citep[solid lines;][Yung et al., in prep]{behroozi15,mason15,wilkins22,dayal17} span a wide range, reflecting the lack of observational constraints prior to {\it JWST}. Across the full NGDEEP NIRCam mosaic (to the F277W 5$\sigma$ limit) NGDEEP should cover $\sim$50--1100 galaxies at $z\sim10$, 3--25 galaxies a $z \sim$ 13, and up to 10 galaxies at $z =$ 13.5-15, distinguishing between these models. We note that early results from CEERS \citep{finkelstein23} are consistent with the most optimistic of these models \citep{behroozi15}.
\textit{Right:} Model uncertainties highlight our state of knowledge of important physical processes. Purple shading shows the plausible range of the dependence of SFR on gas surface density, which primarily affects the bright-end which is constrained by wider-area programs like CEERS and COSMOS-Web \citep{casey23}. Red shading denotes the plausible range in stellar feedback mass loading factors, which primarily affect the faint end \citep{yung19}. The symbols show the precision achievable by CEERS and NGDEEP+JADES (assuming the fiducial \citealt{yung19} luminosity function; dotted line). At $M_{\mathrm{UV}}=-17.5$ (NGDEEP-NRC $10\sigma$ limit), NGDEEP-NRC alone will significantly constrain these models ($\sim$4.5$\sigma$, including cosmic variance, improving to $\sim$5.5$\sigma$ with the eventual combination with JADES), placing the first definitive constraints on feedback processes in early galaxies.
\label{fig:lfs}}
\end{figure*}

\section{NGDEEP-NRC Science Cases: A Public and Parallel Deep Survey Towards First Light}\label{sec:nrc_science}

NGDEEP parallel NIRCam imaging will achieve m$_\mathrm{AB, 5\sigma}=30.7$ (30.4) from 1--5$\mu$m over 5 (11) arcmin$^2$, on the deepest F814W imaging in the sky (m$_{AB, 5\sigma}=30.0$). These data will allow some of the first robust constraints on galaxy formation at $z>10$. Public NGDEEP data will enable immediate community-led deep-field legacy science up to magnitudes comparable or deeper than proprietary GTO NIRCam surveys, all coming in parallel at essentially no added observational cost.

\subsection{Physical Processes Regulating the Emergence of the First Galaxies}\label{sec:first_galaxies}

Over the first 500 Myr of cosmic time ($z\!\!>$10) galaxies began to coalesce, enriching and ionizing their environments, altering  subsequent gas accretion and star formation.  {\it HST} has only scratched the surface with just a handful of $z\sim$9--11 (mostly) tenuous candidates \citep{mcleod15,oesch18,bouwens19,bagley22,finkelstein22}.  Prior to {\it JWST}, the number of faint galaxies ($M_{UV}{<}-$17) found by different surveys was in tension, and there is evidence for both smooth or accelerated decline in the observable cosmic SFR density  \citep{finkelstein15,bouwens15,oesch18,mcleod15,finkelstein16,bouwens19}.  This leaves 
theoretical models highly unconstrained, where differences in star-formation and feedback prescriptions result in a wide range of predictions \citep[Figure~\ref{fig:lfs};][]{behroozi15,mason15,wilkins22,dayal17,behroozi20,yung19,mason15b,wilkins17,gnedin16,dave19}.  This leads to  uncertainties on the details of the history of reionization of the IGM \citep{robertson15,finkelstein19, yung20a, yung20b}.

NGDEEP will provide robust observations of the number density of galaxies at $z > 10$.  This measures the shape of the rest-UV luminosity function (UVLF), 
which constrains the relative strength of processes governing gas conversion
into stars (and these depend on gas density, metallicity, magnetic field strength, turbulence, and feedback mechanism).  Different theoretical predictions for the UVLF at $z>10$ vary by orders of magnitude in galaxy number density \citep[e.g.,][]{behroozi15,mason15,wilkins22,dayal17}, because of differences in adopted sub-grid physics.  Two physical processes are most important for the shape and normalization of the UVLF at high redshift: the star-formation law (i.e., the star formation efficiency as a function of cold gas surface density), and the prescription for stellar feedback.

Using a semi-analytic model (SAM), \citet{yung19} explored how modifications to these relationships alter the UVLF, shown in the right-hand panel of Figure 7.  Altering the star-formation efficiency primarily 
changes the bright end of the UVLF, which will be constrained by medium-depth, wider-field {\it JWST} programs (such as CEERS, \citealt{finkelstein22c}, and COSMOS-Web, \citealt{casey23}).  However, stellar feedback dominates the
UVLF of faint galaxies (as feedback regulates star-formation in lower-mass halos). 
In particular, this feedback depends on the mass-loading factor of ejected gas, and changing this within the currently allowable parameter space changes the number density of the lowest-luminosity galaxies ($M_{UV}{\lesssim}{-18}$) by up to 1~dex.  
NGDEEP provides the missing constraint:  Figure~\ref{fig:lfs} shows that adding NGDEEP observations achieves an uncertainty on the number density of faint galaxies $\sim$4--5$\times$ less than the range currently spanned by models (this includes both counting and a 30\% fractional cosmic variance uncertainty; \citealt{bhowmick20}).  

To make this measurement as accurate as possible requires deep surveys of blank fields. 
This is complementary to surveys of regions that are strongly lensed by clusters of galaxies. Those observations can reach fainter intrinsic luminosities over small volumes, albeit with uncertainties associated with the magnification.
By directly identifying galaxies to $m{>}$30.5~mag, NGDEEP will measure the evolution of the UVLF and SFR density to $z{>}10$--15, and provide unique constraints on stellar feedback physics at $z{>}10$.  In Figure~\ref{fig:lfs} we also list predicted numbers of high-redshift galaxies to be discovered by NGDEEP based on a range of recent models.  As early {\it JWST} results are finding observed yields at the high end of predictions, these numbers may be lower limits \citep[e.g.,][]{castellano22,naidu22,finkelstein22c,finkelstein23,donnan22,harikane22,adams22,atek23,perez-gonzalez23}.

\subsection{The Onset of Chemical Enrichment}\label{sec:enrichment}
As we explore higher redshifts, we will eventually witness the periods during which galaxies have formed no more than a few generations of stars, characterized by extremely low metallicities.  Exactly how low is critical — if all dense gas in the universe is rapidly enriched beyond the critical metallicity ($\sim$10$^{-4}$ Z\sol), both the stellar initial mass function and stellar photospheric temperatures will likely not be dramatically different than those seen in low-metallicity environments.  If the opposite is true, and fairly massive metal-free stars can form down to even $z \sim$ 10, we expect markedly harder stellar spectra, with consequences on the ability of stellar light to reionize the IGM.

Work using {\it HST} data found that the colors of the most distant galaxies are consistent with low (but non-zero) metallicities, without much dust obscuration in the lowest mass galaxies \citep{finkelstein12a,dunlop13,bouwens14,wilkins15}.  NGDEEP NIRCam imaging will push this analysis to $z>10$, allowing us to measure the UV spectral slope $\beta$; where f$_{\lambda}\propto \lambda^{\beta}$ \citep{calzetti94} with four (three) rest-UV colors for galaxies at $z\sim10$ (12).
Simulations at our proposed depths show that we can recover $\beta$ with minimal bias and $\sigma_{\beta}=0.2$ to $z > 13$.  NGDEEP will also improve measures of $\beta$ in galaxies at $z=6-8$ (currently restricted to just one or two colors), measuring $\beta$ with five colors. 

\subsection{The Sites of Early Black Hole Formation}\label{sec:blackholes}
Massive black hole seeds forming at $z > 12$ via direct collapse,  with masses $\sim 10^4 - 10^5\, M_{\odot}$, are expected to 
form in the satellite halos of early star-forming galaxies, that will eventually merge and acquire a stellar component \citep{Agarwal+2016}. 
In these galaxies with overly-massive black holes (referred to as OBGs), the accretion luminosity outshines the stellar component \citep{Natarajan+2017}, offering a unique way to discriminate between light and massive initial black hole seeds. Computing the multi-wavelength energy output of OBGs, they should stand out from typical galaxies via their steep 1--3$\mu$m SED, identified via NGDEEP colors, with candidates cross-correlated with the very deep {\it Chandra} X-ray data in this field.
The boosted infra-red luminosity of OBGs (predicted $m_{AB} < 25$) makes them easily and unambiguously detectable in the proposed NGDEEP survey.  Recent models predict $\sim$2--5 OBG candidates between $z\sim 9-12$ in the proposed NGDEEP fields, revealing the sites of early BH formation and enabling discrimination between early BH seeding models \citep{Ricarte&Natarajan2018}. 

\subsection{Galaxy Morphologies}\label{sec:morphologies}
Ultra-deep sub-kpc imaging across the near-IR is needed to understand how galaxies assembled. The suppression of runaway star formation by feedback is encoded in galaxies' structural and mass-assembly histories. Additionally, models predict various pathways for emergence of galaxy structure at early times \citep[e.g.,][]{wellons2016} and the role of galaxy mergers in the mass assembly of galaxies over the age of the universe remains uncertain. This is especially true for galaxies at $z \gg 2$, at $M_* < 10^{10} M_{\odot}$, and for minor mergers \citep{kaviraj2014, martin2017, mantha2018, duncan2019}. By characterizing structures in and around low-mass galaxies, we can constrain the link between feedback physics and galaxy structure over cosmic time \citep[e.g.,][]{moody2014, oklopcic2017, zhang2019}. Early morphological studies with \textit{JWST} have demonstrated its power to probe the detailed morphologies of galaxies out to very high redshift, to reveal low-surface brightness features in galaxies that were previously undetected with \textit{HST}, and to identify complex structures even at $z>7$ \citep[e.g.,][]{robertson2022, ferreira2022, kartaltepe2022, nelson2022, bowler2022, finkelstein23,treu2023,chen2023}. 

NGDEEP will push morphological analyses into new frontiers through the measurement of rest-frame optical morphologies and sizes of galaxies at very high redshift and those with low-mass, including the structure of spheroids, disks, and clumps  and will characterize the detailed morphological substructure of faint disks, streams, and other low surface brightness features. The deep imaging will enable the detection of galaxies in the process of merging as well as post-mergers through the identification of shells, tidal tails, double nuclei, and other surrounding debris \citep[e.g.,][]{mantha2019, hsiao2022, kokorev2023}. NGDEEP imaging will also push spatially-resolved SED fitting to $z>4$, probing inside-out quenching and the formation of early bulges, by allowing reconstruction of the radial profiles of SFR, stellar mass surface density, and specific SFR of galaxies \citep[e.g.,][]{abdurrouf2023}.


\begin{figure*}
\epsscale{1.2}
\plotone{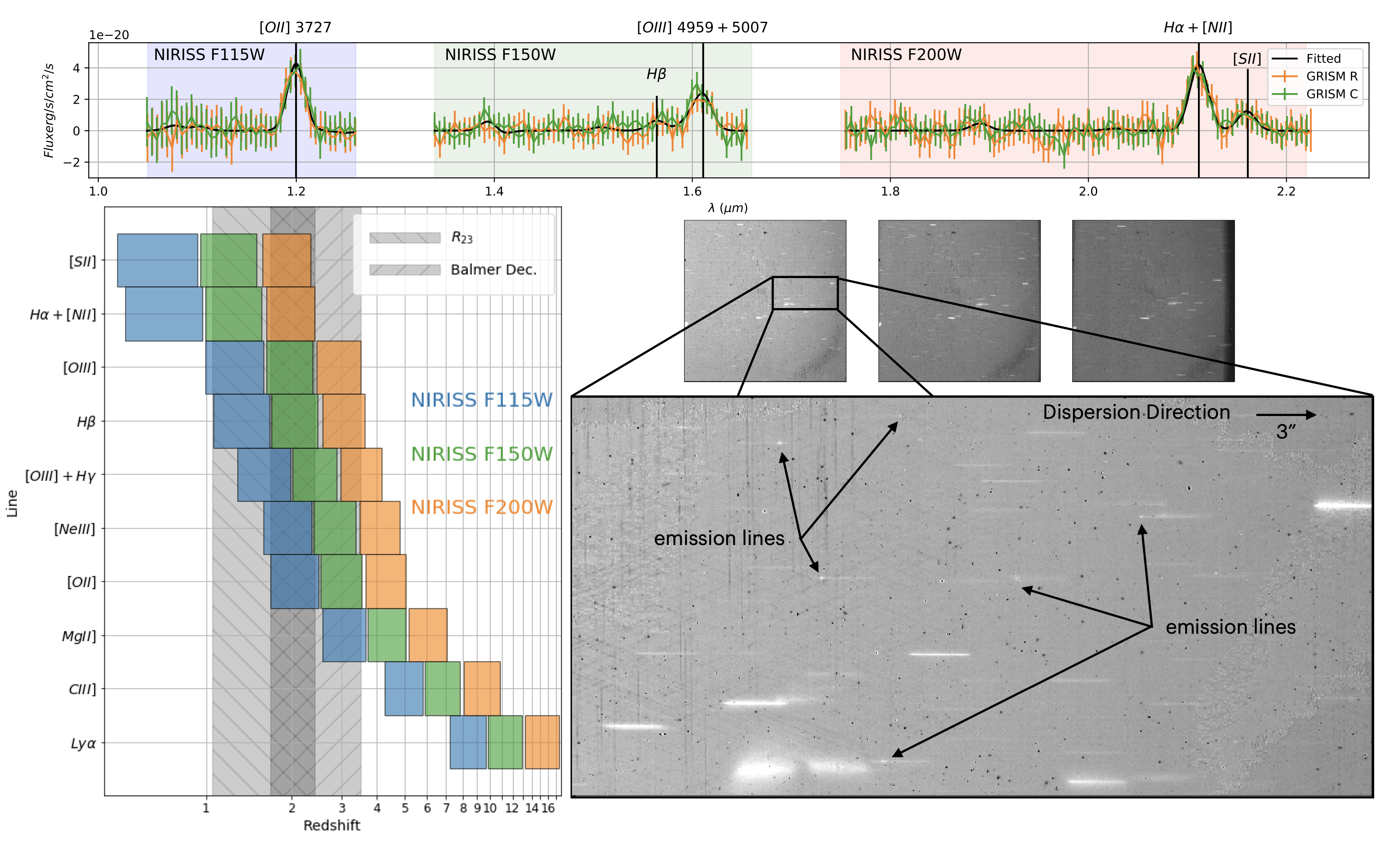}
\caption{NGDEEP-NIS WFSS Simulations. \textit{Bottom Left:} The NIRISS grism covers a wide range of emission lines over a wide range of redshifts. 
\textit{Bottom Right:} Zoomed-in simulated WFSS observations using the F115W cross filter where emission lines are clearly visible. The insets show full images and the region in F115W that is zoomed-in. All known instrumental and WFSS related effects are included in the simulated data. 
\textit{Top:} The GRISMR (orange) and GRISMC (green) extracted NIRISS spectra of a galaxy at $z=2.21$ with $F(H\alpha) = 10^{-17}$ erg/s/cm$^2$ (SNR=22), $10^{7.7}$ $m_{\odot}$. We will measure $\sigma>5$ emission lines in over 1500 galaxies in the HUDF.
\label{fig:nis_sims}}
\end{figure*}

\subsection{Closing the Gap between the \\ Near Field and Deep Fields}
Resolved stellar population studies of Local Group galaxies can be used to reconstruct the high-redshift luminosity function out to $z\sim4$--5 down to very faint magnitudes \citep[e.g.,][]{Weisz14}. The unprecedented depth of NGDEEP's multi-band imaging, combined with deep {\it HST} imaging, will allow us to probe very faint galaxies at $z \sim4$--5 \emph{directly}, and test for consistency with the fossil record.
NGDEEP will probe  to $M_{\rm UV} \sim -15.5$ ($-15.7$) at $z=4$ ($z=5$), overlapping the brightest bins from the \citet{Weisz14} Local Group analysis. Future resolved stellar population studies with {\it JWST} and ground-based ELTs will be able to probe more distant galaxies beyond the Local Group, improving the fossil constraints and the overlap with the NGDEEP constraints.

\section{NGDEEP Simulated Observations}\label{sec:simulations}
We performed a careful evaluation of the feasibility of NGDEEP using a complete set of end to end simulations for both the primary and parallel observations. We used the 
Multi-Instrument Ramp Generator\footnote{\url{mirage-data-simulator.readthedocs.io}} (MIRAGE) to generate mock NGDEEP observations for both \JWST\ instruments. The simulations are based on the NGDEEP APT file, and so reproduce the filters, grisms, exposure specifications, dither patterns and position angles of the planned observations. These simulations include realistic noise, sky background structure, and all known instrument-dependent (e.g., bad pixels, cross talk, etc.) and scene-dependent (e.g., variation in dispersed background, object morphology, crowding) effects for the NIRISS WFSS and NIRCam imaging. We reduced and analyzed these data in exactly the same way we plan to do with using real data. We describe each set of simulations in the following sections. 

\begin{figure*}
\epsscale{1.1}
\plotone{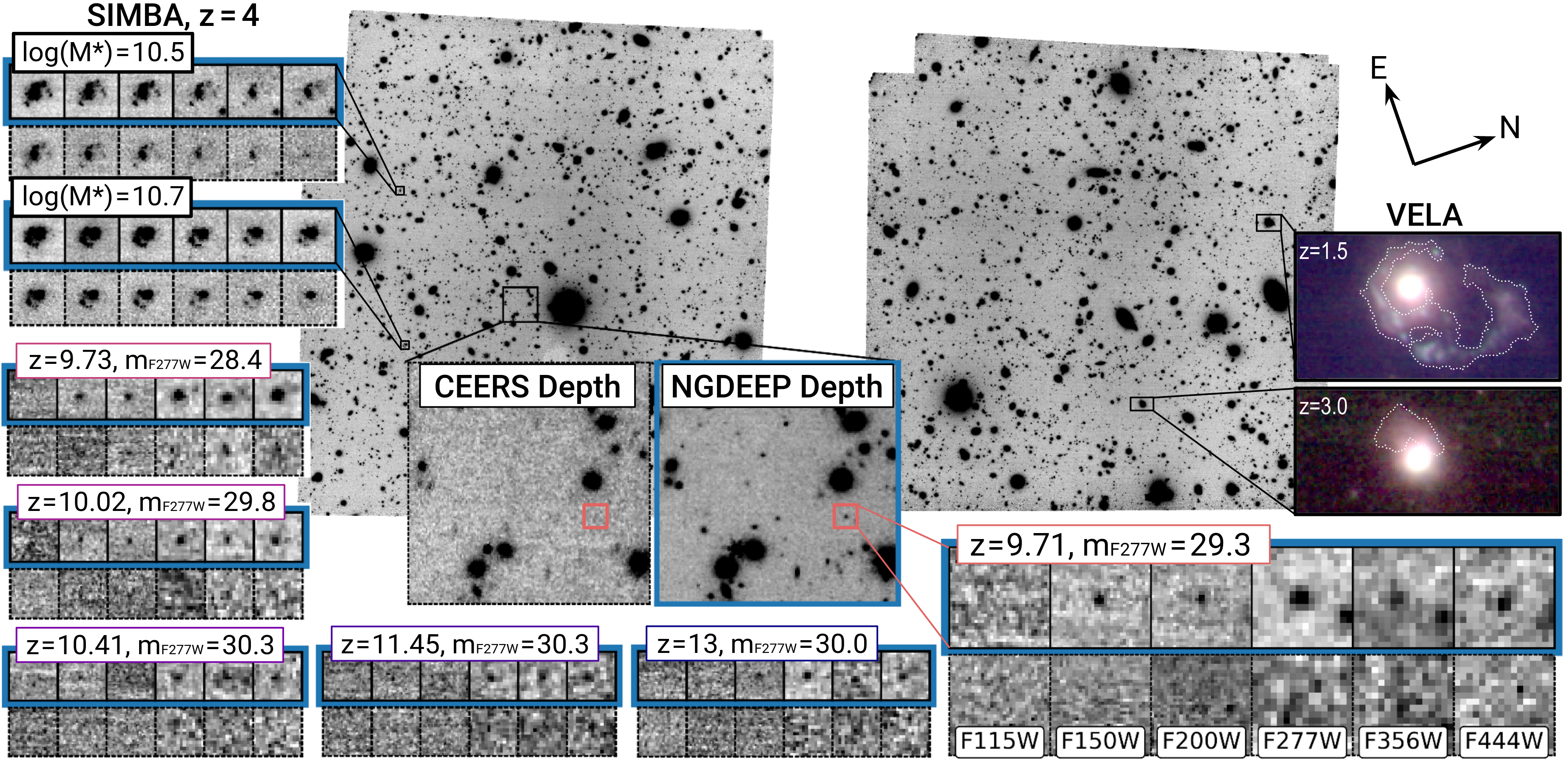}
\caption{Full simulated NGDEEP-NRC F277W 22.3 hour mosaic, displaying the footprint of the proposed observations, including all instrumental effects and sensitivities. The main input for this simulation is the ultra-deep Santa Cruz SAM lightcone \citep{yung22}. The central inset zooms in on a 10.7\arcsec\ region, demonstrating the increase in depth compared with CEERS. 
On the right, we show composite NGDEEP-NRC images of $z=1.5$ and $z=3$ galaxies from the VELA simulation \citep{simons2019} with faint tidal features (indicated by white dashed lines) clearly detected in the planned exposure times. By resolving low-mass galaxies at early times, NGDEEP will probe a new frontier in galaxy assembly.
On the upper left we show 2\arcsec\ postage stamps of two massive galaxies at $z=4$ generated from the SIMBA simulation \citep{dave19}, with the NGDEEP-NRC images on top and stamps of the same sources imaged at the CEERS depth on the bottom. Extended emission and complicated morphologies are evident in all filters at the NGDEEP-NRC depth.  
We also show 1\arcsec\ postage stamps of six $z>9.5$ galaxies, again compared with CEERS imaging, highlighting the power of NGDEEP-NRC imaging to detect $m>30$ galaxies to $z=13$.
\label{fig:nrc_sims}}
\end{figure*}

\subsection{NGDEEP-NIS Simulations}\label{sec:nis_sims}

As input to the NIRISS WFSS simulations we used the JADES catalog \citep{williams2018}, which provides morphological and spectroscopic data. The resulting input catalogs and spectra were used in conjunction with an APT observing plan and the simulation software MIRAGE  to produce individual uncalibrated NIRISS imaging and WFSS exposures. We therefore produced a set of rate files with dithers and readout mode that were the same as our proposed observations. Spectra from the WFSS rate files were then extracted using the SBE method, following closely the methodology described in \citet{pirzkal2018}. Combined GRISMR and GIRSMC 1D spectra, wavelength and flux calibrated were produced and served to check our expected S/N and we stress that our simulations include as detailed as possible treatment of all the known instrumental effect and of the observing strategy. We show an example of our simulated WFSS observations in Figure~\ref{fig:nis_sims}. This figure also shows the extracted spectra for a source as seen by the GRISMR and GRISMC grisms.


\subsection{NGDEEP-NRC Simulations}\label{sec:nrc_sims}

For the mock NIRCam observation inputs, we used a modified version of the \textit{ultra-deep} simulated lightcone presented by \citet{yung22}. The galaxies therein are simulated using the Santa Cruz semi-analytic model \citep[SAM][]{somerville15b,somerville21,yung19,yung2019b}, with dark matter halos extracted from the IllustrisTNG-100 dark matter-only simulation \citep{nelson19} and Monte Carlo merger trees constructed based on the extended Press-Schechter formalism \citep[e.g.,][]{somerville99, somerville08}. The SAM incorporates the evolution of a variety of physical processes, such as cosmological accretion, cooling, star formation, chemical enrichment, and stellar and AGN feedback. See \citet{yung22} for a concise summary of the internal workflow of the SAM and lightcone construction. The free parameters in the SAM are calibrated to a subset of observed constraints at $z\sim0$ and the model performance at high redshift has been tested extensively and shown to well-reproduce a wide variety of observed constraints \citep{yung19, yung20b, yung21}. 

The predicted star formation and chemical enrichment histories of each mock galaxy are coupled with synthetic stellar SEDs from \citet{bruzual03}, and are forward modelled into rest-frame and observed-frame photometry in the NIRCam filters, accounting for ISM dust \citep{calzetti00} and IGM extinction \citep{madau96}. Each synthetic SED also includes nebular emission lines simulated based on the predicted contributions of young stars, AGN, and post-ABG stellar populations \citep[][Yung, Hirschmann, Somerville et al., in prep.]{hirschmann17,hirschmann19,hirschmann22}. The mock catalog contains galaxies out to $z\sim12$ and is complete to $m_\text{F200W} \sim 34$. The SAM mock galaxies also provide simulated S{\'e}rsic profiles \citep{sersic63,sersic68} with S{\'e}rsic indices and effective radii determined as described in \citet{brennan15}. 

We supplemented this galaxy mock catalog with three additions. First, we injected $\sim$20 galaxies at redshifts $11-15$, with NIRCam filter magnitudes pulled from the SAM extended Press-Schechter catalogs, providing us with realistic photometry for these high-redshift galaxies. We placed these additional sources randomly in the NGDEEP-NRC footprint with the goal of evaluating the expected recovery rate of galaxies with $m\gtrsim30$ and $z>12$, should they exist in the real field. Next, we included postage stamps of extended sources with fully realistic morphologies from both the VELA\footnote{\url{https://archive.stsci.edu/prepds/vela/}}  \citep{simons2019} and SIMBA \citep{dave19} simulations. The VELA simulated sources were included to test the recovery of faint, extended structures such as streams and tidal features, and the SIMBA simulated sources provided tests of morphological measurements of substructures such as spheroids, disks and clumps. 

The full NGDEEP-NRC simulation includes 1400 simulated images (140 exposures each with ten NIRCam detectors). 
We created raw images with MIRAGE and reduced them using the \JWST\ Calibration Pipeline\footnote{\url{jwst-pipeline.readthedocs.io}}. We show an example of our simulated NIRCam observations in Figure~\ref{fig:nrc_sims}, which highlight the ability of these data to identify very high-redshift galaxies fainter than 30th magnitude.

\section{Data Release Plans}\label{sec:datareleases}
The NGDEEP survey builds on {\it HST}'s legacy surveys -- the HDF, UDF, GOODS \citep{giavalisco02}, CANDELS \citep{grogin11,koekemoer11}, etc. -- whose success resulted in part from the immediate release of the data.  NGDEEP also achieves a primary  goal of \JWST: ``First Light and Reionization \ldots to see the first stars and galaxies \ldots", which requires deeper surveys than the ERS programs.  NGDEEP leverages {\it HST}'s legacy by combining \textit{JWST} data with the deepest \textit{HST} imaging in the $B$- and $I$-bands, and will enable the study of galaxies from $z{\sim}$1--2 to $z{>}10$.
Progressing from having a few secure $z{>}$10 candidates to having a robust sample of galaxies will require a concerted effort over the first several years of {\it JWST} operations.   NGDEEP will form a crucial cornerstone of Cycle 1 extragalactic surveys, informing the design of future {\it JWST} programs (e.g., deep field proposal strategies in Cycle 3 and beyond).  With NGDEEP and GTO programs (such as JADES) as building blocks, we envision programs executed by multiple teams to ultimately push several fields to ambitious depths early in the mission and get data to the community rapidly. NGDEEP offers a very efficient first step toward this goal.

To that end, we will rapidly deliver high--quality data products to enable broad community science from NGDEEP. 
We will reduce the NGDEEP-NRC observations using the STScI \JWST\ Calibration Pipeline with custom additions and modifications we have identified through our work with CEERS imaging \citep{bagley22b}. These modifications include the removal of low-level features such as wisps and $1/f$ correlated noise, a careful astrometric alignment, and a global background subtraction, steps that are necessary to achieve the expected NGDEEP-NRC imaging depths.  
The NGDEEP-NIS observations will be processed through Stage 1 of the \JWST\ Calibration Pipeline with any necessary modifications, and the spectra will be extracted using the SBE method \citep{pirzkal17}.
The NIRISS release will include 2D and 1D calibrated spectra of each source detected in the field with an accurate correction for spectral contamination as demonstrated by \citet{pirzkal17}. 
Our data-management plan is built on successful {\it Hubble}
Treasury programs, where speed is valued over perfection for the first v0.5 releases, with limitations noted in the accompanying documentation. Our first reductions will therefore represent a best effort based on the current state of the instrument calibrations, especially that of the NIRISS WFSS mode. Later v1 releases will include more extensive testing and refinement, and be accompanied by well-tested catalogs.
 Our released catalogs will be accompanied by Python Jupyter notebooks to train  users to interact with data products.  All  products will be shared with STScI, and hosted on our own team website. Table 3 shows the NGDEEP data release schedule.

\begin{table*}
\centering
\caption{NGDEEP Data Release Plan} \label{tab:datareleases}
\begin{tabular}{m{0.35\linewidth} | cc | m{0.4\linewidth}}
\hline
\hline
Data Product & v0.5 & v1 & Notes \\
\hline
%
Reprocessed NIRISS images & 3 & 6 & Anticipated improvements: registration, cosmic ray rejection, background estimation \\[2mm]
Extracted NIRISS spectra & 3 & 9 & 1 dimensional spectra and contamination maps \\[2mm]
NIRISS emission line catalog and redshifts & 6 & 12  & Emission--line analysis \\[2mm]
NIRISS+HST Catalog & \nodata  & 12 & Continuum + emission--line analysis \\
\hline
Reprocessed NIRCam images & 3 & 6 & Anticipated improvements: registration, cosmic ray rejection, background estimation \\[2mm]
NIRCam+HST photometric catalog & 3 & 9 & PSF--matched photometry across all bands (v1=NIRCam, v2 includes \emph{HST}) \\[2mm]
NIRCam+HST morphology & \nodata & 12 & Sersic fits + non-parametric morphologies \\
\hline
\end{tabular}
\begin{tabular}{l}
The version columns refer to the months following acquisition when each data product will be released.  \\
These are estimates, with the exact releases depending on data reduction and analysis timescales. \\
\end{tabular}
\end{table*}

\section{Summary}\label{sec:summary}

We present the Next Generation Deep Extragalactic Exploratory Public Survey, which is obtaining deep NIRISS WFSS of the HUDF and deep NIRCam imaging in the HUDF-Par2 parallel field. 
The NGDEEP observations are split across two position angles, with V3\_PA=70$^{\circ}$ observed in February 2023 (Epoch 1) and V3\_PA=67$^{\circ}$ scheduled for January 2024 (Epoch 2). The position angles are chosen to improve the NIRISS emission line identification and contamination modeling while maximizing the overlap between the two NIRCam imaging sets.

The NGDEEP NIRISS observations (NGDEEP-NIS) are obtaining spectroscopy from $1-2.2$\micron\ with the GRISMR and GRISMC grisms dispersed through the F115W, F150W and F200W filters. The integration times are distributed to achieve approximately uniform emission line sensitivities at all wavelengths, with expected 5$\sigma$ integrated line sensitivities of 1.2, 1.3 and $1.5\times 10^{-18}$ ergs s$^{-1}$ cm$^{-2}$ for F115W, F150W, and F200W, respectively. The NGDEEP-NIS observations will detect emission lines from $>$1000 galaxies, many of which have multiple lines.
The unbiased nature of the NGDEEP-NIS slitless spectroscopy will result in near infrared spectroscopic measurements for every source in the HUDF, complemented by the deep ACS F435W imaging in the field. 

The NGDEEP NIRCam observations (NGDEEP-NRC) are obtaining imaging from $1-5$\micron\ with the F115W, F150W, F200W, F277W, F356W, and F444W filters. The NIRCam imaging will reach comparable depths in all filters ($m=30.7-30.9$, $5\sigma$ resolved source) and $m=31.2$ in F115W. The NGDEEP-NRC observations are defined by the NIRISS observing strategy, including the exposure times, dither patterns, and NIRISS direct imaging. 
There are therefore multiple tiers of depth in the NIRCam imaging, with $\sim$5 arcmin$^2$ of the deepest imaging where the two position angles overlap. 
This imaging will enable the discovery of $\sim$30--100 galaxies at $z\gtrsim11$, probing the faint end of the rest-UV luminosity function 2 magnitudes fainter than that possible with CEERS. NGDEEP-NRC is supplemented by deep ACS F814W imaging covering the NIRCam footprint, with 50\% of the footprint covered by the deepest F814W imaging on the sky (m$_{5\sigma} \approx$ 30).

Together, these coordinated parallel observations are designed to explore the dominant feedback mechanisms in low-mass galaxies across cosmic time. NGDEEP will constrain:
\begin{itemize}
\item the redshift evolution and scatter in the MZR slope from $z \sim$ 1--5;
\item the stochasticity, burstiness and variability of star formation at $z=0.7-2.3$;
\item the physical processes (stellar feedback and star formation efficiencies) regulating the emergence of the first galaxies;
\item the evolution of chemical enrichment starting as early as $z\gtrsim12$;
\item the sites and mechanisms of early black hole formation;
\item the link between feedback and morphological structures in and around low-mass galaxies.
\end{itemize}
As a program continuing the legacy of \hst\ ultra deep field astronomy, NGDEEP will enable many additional community-led explorations into the study of galaxies from cosmic noon ($z\sim1-2$) to the early epochs of galaxy formation ($z>10$).  

NGDEEP provides the deepest publicly available spectroscopy and imaging obtained in Cycle 1, making it one of the first \JWST\ deep fields. When combined with NGDEEP Epoch 2 as well as the JADES and MIRI GTO programs, these legacy observations will transform deep field science and dramatically enhance our understanding of the universe.

\begin{acknowledgements}
MBB, SLF, DAB, GCKL, CMC, OACO, KC, SF, and RLL acknowledge that the location where part of this work took place, the University of Texas at Austin, that sits on indigenous land. The Tonkawa lived in
central Texas and the Comanche and Apache moved through this area. We
pay our respects to all the American Indian and Indigenous Peoples and
communities who have been or have become a part of these lands and territories
in Texas, on this piece of Turtle Island.

We thank the entire \JWST\ team, including the engineers for making possible
this wonderful over-performing telescope, the instrument and commissioning teams for establishing and characterizing observatory performance, the scheduling team for squeezing in Epoch 1 of NGDEEP as the observing window was closing, and the pipeline teams for their work over the years building and supporting the pipeline.
The authors acknowledge the Texas Advanced Computing Center (TACC) at The
University of Texas at Austin for providing HPC and visualization resources
that have contributed to the research results reported within this paper.
This work is based on observations with the NASA/ESA/CSA \JWST\ obtained from the Mikulski Archive for Space Telescopes at the Space Telescope Science Institute (STScI), which is operated by the Association of Universities for Research in Astronomy (AURA), Incorporated, under NASA contract NAS5-03127.

We acknowledge funding support from STScI through JWST-GO-2079.
PGP-G acknowledges support  from  Spanish  Ministerio  de  Ciencia e Innovaci\'on MCIN/AEI/10.13039/501100011033 through grant PGC2018-093499-B-I00.
\end{acknowledgements}

\vspace{5mm}
\facilities{\JWST\ (NIRCam, NIRISS), \HST\ (ACS, WFC3)}

\software{Astropy \citep{astropy},
          Drizzle \citep{fruchter2002},
          \JWST\ Exposure Time Calculator (\url{jwst.etc.stsci.edu}),
          MIRAGE (\url{mirage-data-simulator.readthedocs.io}),
          SciPy \citep{virtanen2020},
          STScI \JWST\ Calibration Pipeline (\url{jwst-pipeline.readthedocs.io})}


\bibliographystyle{aasjournal}
\bibliography{wdfbib}






\end{document}